\documentclass[aps,pra, reprint,superscriptaddress,a4paper,longbibliography]{revtex4-1} 
\usepackage{amsmath,amssymb} 
\usepackage[utf8]{inputenc} 
\usepackage{graphicx} 
\usepackage[T1]{fontenc} 
\usepackage{hyperref}
\hypersetup{colorlinks=true,citecolor={black},linkcolor={black},urlcolor={blue}} 
\usepackage{xr} 
\usepackage{xcolor}
\usepackage{textcomp}
\usepackage{paralist}
\usepackage{bm}

\usepackage[separate-uncertainty=true,multi-part-units=single,group-separator={,}]{siunitx} 
\DeclareSIUnit[per-mode=symbol,per-symbol=p]{\ueV}{\micro\electronvolt}

\newcommand{\mi}{\mathrm{i}}

\newcommand{\ve}{\varepsilon}
\newcommand{\half}{\textstyle{\frac{1}{2}}}
\newcommand{\ihalf}{\textstyle{\frac{\mathrm{i}}{2}}}

\begin{document}

\title{Spontaneous spin bifurcations and ferromagnetic phase transitions in a spinor exciton-polariton condensate}

\author{H. Ohadi}
\email{ho278@cam.ac.uk}
\affiliation{ Department of Physics, Cavendish Laboratory,
    University of Cambridge, Cambridge CB3 0HE, United Kingdom }
\author{A. Dreismann}\affiliation{ Department of Physics, Cavendish Laboratory, University of Cambridge, Cambridge CB3 0HE, United Kingdom } 
\author{Y. G. Rubo} \affiliation{
   Instituto de Energías Renovables, Universidad Nacional Autónoma de México,
   Temixco, Morelos, 62580, Mexico} 
\author{F. Pinsker} \affiliation{ Department of Applied Mathematics and
    Theoretical Physics, University of Cambridge, Cambridge CB3 0WA, United
    Kingdom } 
\affiliation{Clarendon Laboratory, University of Oxford, Parks Road, Oxford OX1
    3PU, United Kingdom}
\author{Y. del Valle-Inclan Redondo}\affiliation{ Department of Physics, Cavendish Laboratory, University of Cambridge, Cambridge CB3 0HE, United Kingdom } 
\author{S. I. Tsintzos} \affiliation{ Foundation for Research and Technology–Hellas, Institute of Electronic Structure and Laser, 71110 Heraklion, Crete, Greece }
\author{Z. Hatzopoulos} \affiliation{ Foundation for Research and Technology–Hellas, Institute of Electronic Structure and Laser, 71110 Heraklion, Crete, Greece } \affiliation{ Department of Physics, University of Crete, 71003 Heraklion, Crete, Greece }
\author{P. G. Savvidis}
\affiliation{ Department of Physics, Cavendish Laboratory, University of Cambridge, Cambridge CB3 0HE, United Kingdom }
\affiliation{ Foundation for Research and Technology–Hellas, Institute of Electronic Structure and Laser, 71110 Heraklion, Crete, Greece } 
\affiliation{ Department of Materials Science and Technology, University of Crete, 71003 Heraklion, Crete, Greece }
\author{J. J. Baumberg}
\email{jjb12@cam.ac.uk}
\affiliation{ Department of Physics, Cavendish
    Laboratory, University of Cambridge, Cambridge CB3 0HE, United Kingdom }

\begin{abstract}
We observe a spontaneous parity breaking bifurcation to a ferromagnetic state in
a spatially-trapped exciton-polariton condensate. At a critical bifurcation
density under nonresonant excitation, the whole condensate spontaneously
magnetizes and randomly adopts one of two elliptically-polarized (up to 95\%
circularly-polarized) states with opposite handedness of polarization.  The
magnetized condensate remains stable for many seconds at \SI{5}{K}, but at
higher temperatures it can flip from one magnetic orientation to another. We
optically address these states and demonstrate the inversion of the magnetic
state by resonantly injecting 100-fold weaker pulses of opposite spin.
Theoretically, these phenomena can be well described as spontaneous symmetry
breaking of the spin degree of freedom induced by different loss rates of the
linear polarizations.
\end{abstract}

\maketitle

Condensation of exciton-polaritons (polaritons) spontaneously breaks the global
phase
symmetry~\cite{deng_condensation_2002,kasprzak_bose-einstein_2006,balili_bose-einstein_2007,baumberg_spontaneous_2008,ohadi_spontaneous_2012}.
Owing to their easy optical interrogation, high-speed (ps) interactions, and
macroscopic coherence (over hundreds of microns)~\cite{nelsen_dissipationless_2013}, polariton condensates are excellent
candidates to probe and exploit for
sensing~\cite{sturm_all-optical_2014,dreismann_coupled_2014},
spinoptronics~\cite{amo_exciton-polariton_2010,ballarini_all-optical_2013,cerna_ultrafast_2013},
new optoelectronic
devices~\cite{nguyen_realization_2013,bhattacharya_solid_2013,schneider_electrically_2013},
and quantum simulators~\cite{buluta_quantum_2009}. The driven-dissipative multicomponent polariton system can undergo additional bifurcations and condense into states which are not eigenstates of the single-particle Hamiltonian, but many-body states with reduced symmetry \cite{aleiner_radiative_2012,zhang_weak_2015}. Thus, we should expect that two-component
exciton-polariton condensates can also show spontaneous symmetry breaking
bifurcations in their polarization state. Spin studies of microcavity polaritons
have been of great interest in recent
years~\cite{lagoudakis_stimulated_2002,martin_polarization_2002,kavokin_polarization_2003,kavokin_quantum_2004,renucci_microcavity_2005,krizhanovskii_rotation_2006,gippius_polarization_2007,leyder_observation_2007,paraiso_multistability_2010,gao_polariton_2012,kammann_nonlinear_2012,takemura_polaritonic_2014}.
However, spontaneous symmetry-breaking bifurcation of spin has not been observed
before.

Here, we demonstrate spontaneous magnetization in an exciton-polariton
condensate, as a direct result of bifurcations in the spin degree of freedom.
Utilizing an optically trapped geometry, condensates spontaneously emerge in
either of two discrete spin-polarized states that are stable for many seconds,
$>10^{10}$ longer than their formation time. These states emit highly
circularly-polarized coherent light (up to 95\%) and have opposite circular
polarizations. The condensate stochastically condenses in a left- or right-circularly polarized state, with an occurrence likelihood that can be controlled by the ellipticity of the nonresonant pump. The two spin-polarized states can be initialized and switched
from one state to another with weak resonant optical pulses. Our system has
potential applications in sensing, optical spin memories and spin switches, and
it can be implemented for studying long-range spin interactions in polariton
condensate lattices.

This article is structured as follows: in Section~\ref{sec:intro} we review
trapped polariton condensates and the current understanding of polarization in
untrapped polariton condensates. In Section~\ref{sec:circbuildup} we present the
key theme of this work, which is the spontaneous buildup of stochastic circular
polarization. In Section~\ref{sec:theory} we propose a theoretical framework for
the phenomena discussed in this work. We show that stochastic circular
polarization is a signature of spontaneous parity breaking. In
Section~\ref{sec:res} we time-resolve the coherent driving of the spin, with
resonant excitation. We furthermore investigate the stability of the
spin-polarized states against thermal noise and conclude in
Section~\ref{sec:conc}.

\section{Introduction\label{sec:intro}}

\begin{figure}[htb]
	\centering 
	\includegraphics[width=0.48
    \textwidth]{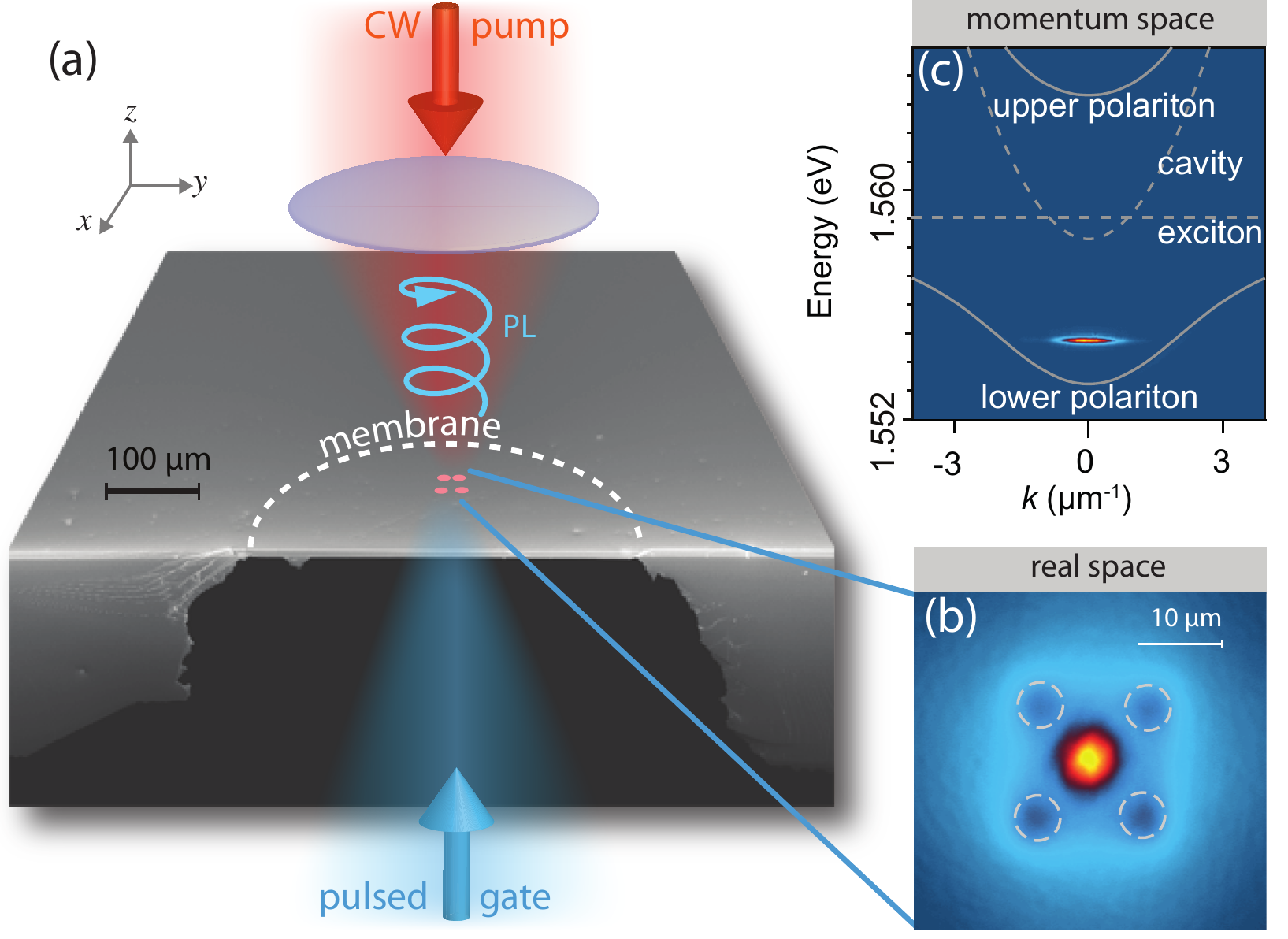} \caption{(a) Scanning electron microscopy image
        of a membrane microcavity, cleaved across the middle. The
        membrane allows resonant excitation from the back side of the cavity.
        (b) Above threshold hot polaritons condense into a single coherent state
        located at the center of the trap (pump spots
        shown by dashed circles), which is found at $k_\parallel=0$ (c). The spin state of the condensate is determined
        by studying the emission polarization.}
    \label{fig:1}
\end{figure}

Exciton-polaritons are spinor particles formed by the strong coupling of
excitons in a semiconductor quantum well with photons in the microcavity in
which they are embedded~\cite{kavokin_microcavities_2007}. We create
optically-trapped polariton condensates by nonresonant excitation of a
semiconductor microcavity membrane~[see FIG.~\ref{fig:1}(a) and
Appendix~\ref{app:expmethods}]. The excitation beam is shaped into a 4-spot
pattern [shown by dashed circles in FIG.~\ref{fig:1}(b)]. The short-wavelength
continuous wave (CW) linearly-polarized pump injects an electron-hole plasma at
each pump spot, which rapidly relaxes to form excitons, in the process losing
all phase information. These reservoir excitons then scatter into polariton
states via multiple phonon-polariton and stimulated polariton-polariton
collisions~\cite{savvidis_angle-resonant_2000}, and they feed the zero-momentum ground
state at the center of the trap. Because of their large effective mass, excitons
typically diffuse only very small distances and stay within \SI{1}{\um} of the
pump spots.  Microcavity polaritons, however, are \SI{10000} times lighter giving
longer diffusion lengths. Driven by their repulsive excitonic interactions,
polaritons can thus travel large distances away from the pump spots within their
lifetime~\cite{wertz_spontaneous_2010}.  Once the density inside the trap
exceeds the condensation threshold, a macroscopically coherent condensate is
formed [FIG.~\ref{fig:1}(c)]. Because the condensate overlaps only weakly with
the pump spots, it shows a narrower linewidth and less decoherence than
unconfined condensates~\cite{askitopoulos_polariton_2013}.  The optical trapping
method used here is similar to optical lattices in cold atomic
systems~\cite{bloch_ultracold_2005}, but with the major difference that the
optical potential also provides gain~\cite{wertz_spontaneous_2010,
wouters_excitations_2007, tosi_sculpting_2012}.

Polaritons in quantum-well microcavities have two $J_z = \pm 1$ (spin-up or
spin-down) projections of their total angular momentum along the growth axis of the
structure, which correspond to right- and left-circularly polarized photons
emitted by the cavity, respectively. When the excitation is linearly-polarized
an equal population of spin-up and spin-down excitons forms in the reservoir.  An
initially spin-balanced reservoir, in the absence of pinning to any
crystallographic axis, is expected to give a condensate with a stochastic linear
polarization~\cite{shelykh_polarization_2006}.  In most experiments with
polariton lasers the condensates have been found to be linearly polarized along
one of the crystallographic
axes~\cite{deng_condensation_2002,kasprzak_bose-einstein_2006}. Nevertheless, in
some cases a circularly polarized polariton lasing has also been
observed~\cite{martin_polarization_2002,
shelykh_semiconductor_2004,krizhanovskii_rotation_2006, roumpos_signature_2009,
ohadi_spontaneous_2012}.  Formation of a circularly polarized condensate is
usually associated with the effects of TE-TM splitting, or bosonic amplification
of the seed polarization of condensates. In all these cases, a circularly
polarized condensate is observed when the symmetry between the spin-up and
spin-down polaritons has been \emph{explicitly} broken in some fashion, by
either the pumping geometry (seeding with a circularly polarized pump) or by the
imposed rotation of the Stokes parameters due to polarization splitting. As a result,
the observed circular polarization was never stochastic (i.e. it is fixed each time
the condensate is excited, but it is different on each realization).

\section{Spontaneous buildup of circular polarization\label{sec:circbuildup}}

\begin{figure*}
	\centering 
	\includegraphics[width=1
	\textwidth]{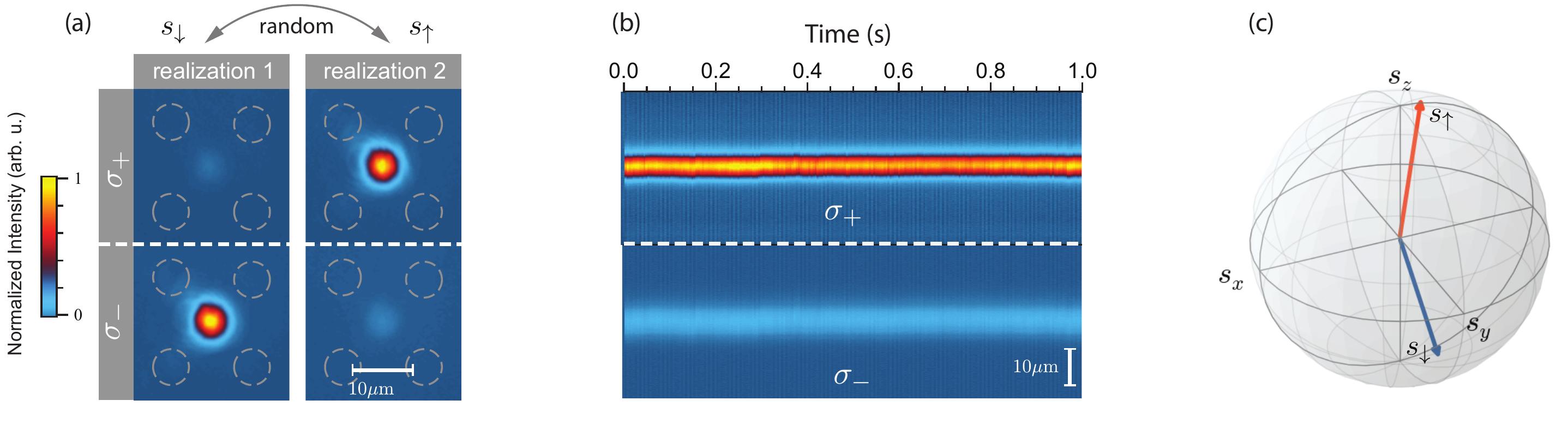} \caption{
       (a) Polarization-resolved spatial image of the only two realizations
       observed in a single trapped condensate in a 4-spot trapping geometry.
       (b) At $T=\SI{5}{\K}$ the spin-polarized states remain stable for many
       seconds. (c) Simultaneously measured components of the pseudospin for
       1000 realizations. The total degree of polarization is $0.93\pm0.03$ and
       the two average pseudospin states are $s_\downarrow=[-0.22, 0.19, -0.94]$
       (blue vector) and $s_\uparrow=[-0.22, -0.14, 0.96]$ (orange vector).  The
       measurement error for each component is $<5\%$, and the variance is
       $\sim1\%$.  
    \label{fig:Random} } 
\end{figure*}

Our experiments reveal a completely different behavior to the previous reports on the polarization of polariton condensates. We observe a strong
degree of circular polarization $60\%<\vert s_z \vert<95\%$
[FIG.~\ref{fig:Random}(a)] when the excitation is linearly polarized (to better
than 1 in $10^{5}$), which is stable for many seconds
[FIG.~\ref{fig:Random}(b)].  The condensate stochastically adopts either of two
opposite circular-polarization states in each realization of the experiment. We
call these two states the spin-up ($s_\uparrow$) and spin-down ($s_\downarrow$) states.  By mapping the
polarization of the photoluminescence (PL) we measure the polarization vector
$s_{x,y,z}=(I_{H,D,\circlearrowright}-I_{{V,A,\circlearrowleft}})/(I_{H,D,\circlearrowright}+I_{{V,A,\circlearrowleft}})$,
where $I$ is the measured intensity for horizontal ($H$), vertical ($V$), diagonal
($D$), anti-diagonal ($A$), right-circular ($\circlearrowright$) and left-circular
($\circlearrowleft$) polarizations. We measure all components of the
polarization vector (pseudospin) simultaneously and plot the mean of the
spin-up and spin down states on the Poincaré sphere separately for 1000
realizations [FIG.~\ref{fig:Random}(c)]. In each realization the wavefunction of
the condensate spontaneously collapses into one of the two discrete
spin-polarized states which have opposing circular and diagonal components,
marked by blue and orange vectors in FIG.~\ref{fig:Random}(c). The linear axis
along which the pseudospin flips (marked here as diagonal) does not depend on
the geometry of the trap, and it changes direction with the position of the
condensate on the sample.

\begin{figure}
	\includegraphics[width=0.48
	\textwidth]{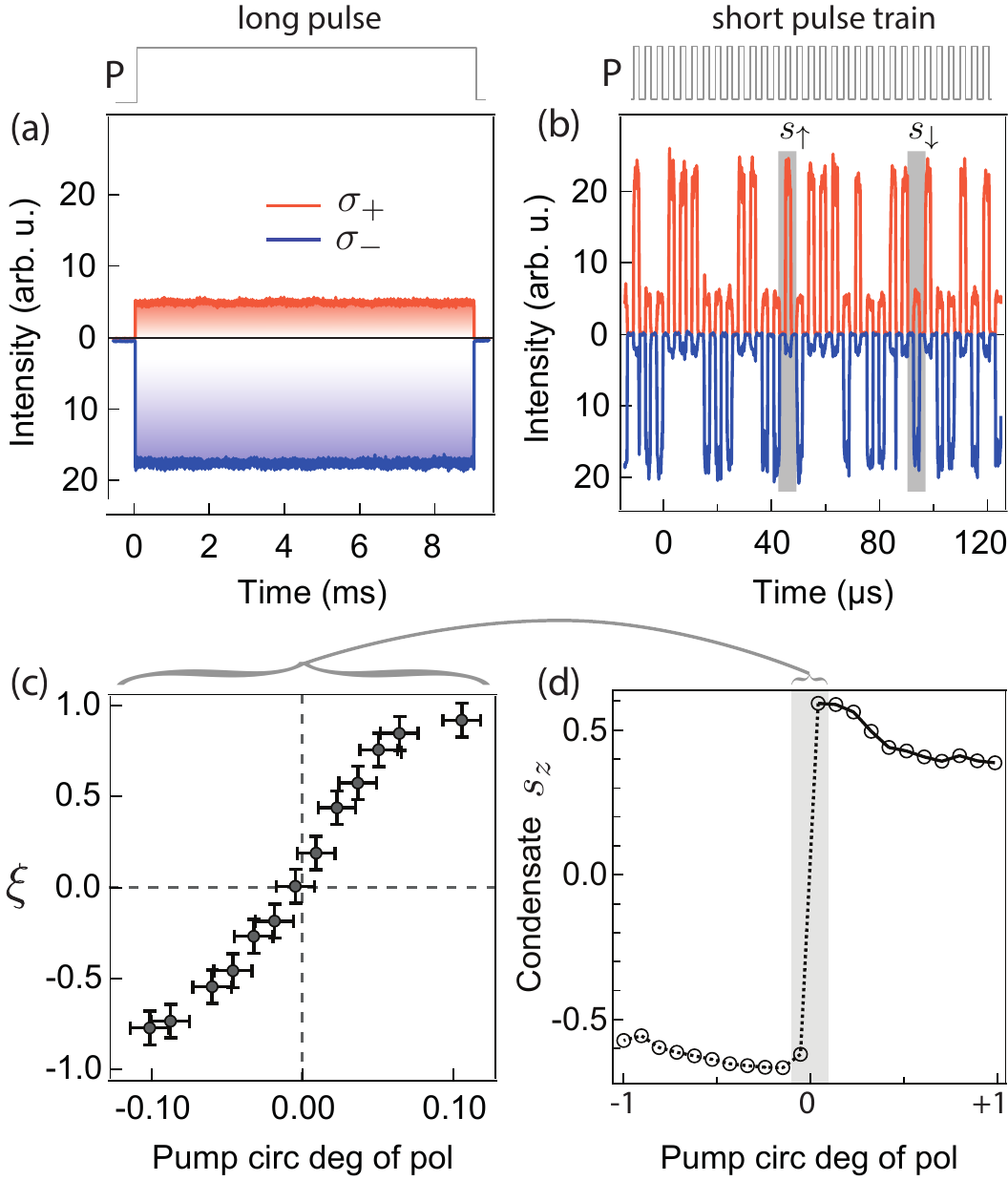} \caption{
        (a) Emitted circular $\sigma_+$ and $\sigma_-$ intensities for a condensate realization when $P>P_c$, created by a long pulse (marked by P). The condensate randomly picks the $s_\downarrow$ state here. Note the opposite axis directions for $\sigma_+$ and $\sigma_-$ components. (b) Same
        as (a), with pump intensity modulated by square wave (shown on top by
        P).  A cut from the middle of a \SI{9}{\milli\second} exposure is shown,
        where each square pulse corresponds to a single realization. The
        condensate randomly picks $s_\uparrow$ or $s_\downarrow$  states (marked
        by grey rectangles).  (c) Occurrence contrast $\xi$ of spin-up and
        spin-down realizations measured in (b) vs the ellipticity of the pump.
        (d) Measured circular polarization of condensate vs pump ellipticity.
        Dashed lines mark the region where the condensate initializes stochastically in
        spin-up or spin-down states in each realization.
    \label{fig:pulsetrain} } 
\end{figure}

To demonstrate that the buildup of circular polarization is truly spontaneous,
we illuminate the sample with a long-duration pulse of \SI{9}{\ms} using an acousto-optic modulator (AOM). A condensate
builds up and picks a random state (e.g. $s_{\downarrow}$) and stays in that
particular state as long as the pump pulse lasts [FIG.~\ref{fig:pulsetrain}(a)]. We then
repeat the same measurement but this time we modulate the pump intensity so that
a condensate is created and destroyed every \SI{2}{\us}.  FIG.~\ref{fig:pulsetrain}(b)
shows how the condensate, although being stable for many seconds, picks a random
polarization at every successive realization due to random initial conditions at
the onset of condensation. In this case the parity symmetry is broken spontaneously. The buildup of circular polarization is independent of the precise position on the sample, of the sample orientation, of the pump spot orientation, and of the polariton detuning.

We can also \emph{explicitly} break the symmetry in our experiments by changing
the ellipticity of the pump laser and measure the contrast of the occurrence
frequency of spin-up ($f_{\uparrow}$) to spin-down ($f_{\downarrow}$)
condensate realizations,
$\xi=(f_{\uparrow}-f_{\downarrow})/(f_{\uparrow}+f_{\downarrow})$.  The
probability of a realization resulting in state $f_{\uparrow\downarrow}$ is
equal to $(1\pm\xi)/2$ respectively. FIG.~\ref{fig:pulsetrain}(c) shows $\xi$
as a function of the pump circular polarization ($s_{z,\text{pump}}$) averaged
over 1000 realizations for each point. As the ellipticity of the pump is
increased from linear ($s_{z,\text{pump}}=0$) to right circular
($s_{z,\text{pump}}>0$) the probability of creating a condensate in the spin-up
state increases; conversely, for the left-circularly polarized pump
($s_{z,\text{pump}}<0$), the probability of creating a spin-down condensate
increases. With a linearly-polarized pump we have an equal probability of
creating a spin-up or spin-down condensate. Although the pump laser is
nonresonant with the final polariton states, the initially created carrier spin
is not entirely randomized during their multiple carrier-carrier and exciton-phonon
scatterings~\cite{roumpos_signature_2009,ohadi_spontaneous_2012,kammann_nonlinear_2012,li_incoherent_2015}. As a result of this incomplete
spin relaxation of the excited carriers, changing the ellipticity of the pump
breaks the symmetry of the condensate toward the same circular polarization as
that of the pump. For pump circular polarizations far greater than 0.1 the
condensate is formed deterministically in the same polarization state as that
of the pump [FIG.~\ref{fig:pulsetrain}(d)].

An interesting question here is why we observe the spontaneous buildup of
circular polarisation, as opposed to the linear polarisation that is widely
reported in the
literature~\cite{deng_condensation_2002,kasprzak_bose-einstein_2006}. The key
difference between the experiments presented in this work and other studies of
polariton condensation lies in the excitation geometry. For our trapped
condensates, the pump and condensate are spatially separated, which critically
reduces the contaminating interactions between the condensate and reservoir. The
large interaction between untrapped condensates and the unpolarized exciton
reservoir results in spin-flip scattering of polaritons with reservoir excitons. If there is a depolarized reservoir on top of the condensate, the spin-flip scattering processes minimize any imbalance between circular components of the condensate. This minimization
leads to quenching of the buildup of circular polarization, forcing the
polaritons to condense only with linear polarization.  Moreover it has been
shown previously that, because of a smaller overlap with the reservoir, trapped
condensates have a smaller linewidth than untrapped
condensates~\cite{askitopoulos_polariton_2013}.  Our careful studies with
temperature and our theoretical calculations (see Section~\ref{sec:thermnoise}), show how spin noise in the system
results in spin flipping of the condensate.
The spin flip rate scales exponentially with noise. Larger linewidth untrapped
condensates have higher spin flip rates, which wash out circular polarization
effects observed here.

\begin{figure}[hbtp]
\centering
\includegraphics[width=0.48\textwidth]{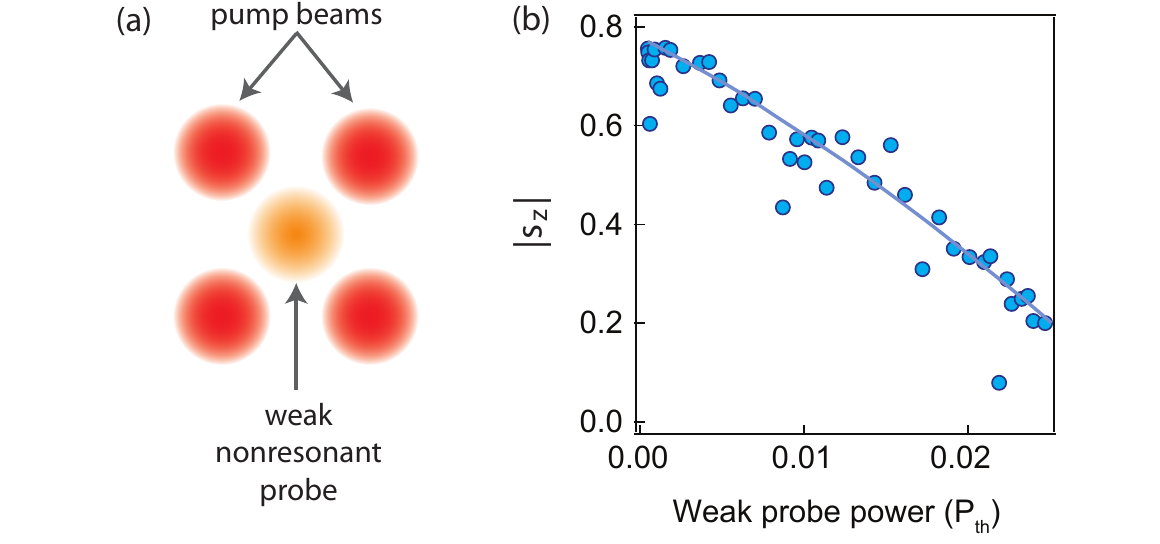}
\caption{(a) Nonresonant linearly polarized weak probe overlapped with the
    condensate. (b) Average circular polarization degree vs power of the probe
    beam. The blue line is a guide to the eye.}
\label{fig:ResOnTop}
\end{figure}

To show the crucial role of the reservoir excitons we place a weak nonresonant
linearly-polarized probe beam on top of the condensate as shown in
FIG.~\ref{fig:ResOnTop}(a). The probe beam which has just a small fraction of
the 4-spot pump power ($<0.025\; P_{th}$), induces a reservoir of excitons that
overlap with the condensate but it does not stop the condensation or reduce the
condensate density below the critical circular polarization density. We
then measure the absolute average circular polarization degree in each
realization, as a function of the weak probe power. As shown in
FIG.~\ref{fig:ResOnTop}(b), the circular polarization degree decreases
monotonously as the probe power increases, demonstrating the quenching of circular
polarization due to the influence of the overlapping reservoir. This measurement
evidences the importance of the separation of the pump-induced
reservoir and the trapped condensate in the observation of ferromagnetic condensates.

Four scenarios might explain the buildup of the stochastic circular polarization
depending on populations of same- and opposite- (or cross-) spin polaritons:
\begin{inparaenum}[(i)]\item a higher interaction energy for cross-spin
     than for same-spin polaritons in a condensate which is in thermal equilibrium,
\item a higher gain from the cross-spin reservoir, \item density-dependent
    losses enhanced by cross-spin condensed polaritons, and \item\label{enum:wl}
    linear polarization energy splitting accompanied by a dissipation rate
    splitting which destabilizes the linearly-polarized condensate (`spin
    bifurcation')\end{inparaenum}. As we will show now, our experimental data
strongly suggest that scenario (\ref{enum:wl}) is the correct explanation.  In
the first scenario (‘energy-minimization’), the condensate free energy is
minimized when it acquires circular polarization.  This situation could happen
in a condensate in thermal equilibrium if the interaction energy of cross-spin polaritons is stronger than that of same-spin polaritons, meaning the coexistence of
cross-polarized polaritons is not favored~\cite{rubo_suppression_2006}.
However, the interaction with opposite spin polaritons is well known to be
weaker than, and opposite to, that of same-spin
polaritons~\cite{vladimirova_polariton-polariton_2010,takemura_polaritonic_2014,takemura_heterodyne_2014}.
Moreover, being externally driven and having short lifetimes, polariton
condensates are generally far from thermal
equilibrium~\cite{kasprzak_formation_2008}. We note that ``energetical'' mechanisms cannot explain why states with elliptical polarization (i.e. not fully circular) are formed. In the second scenario
(‘cross-gain’), in order to acquire a circular degree of polarization, the
condensate must experience larger gain from the cross-spin reservoir than from
the same-spin reservoir. However, the scattering rate from the cross-spin
reservoir into the condensate is measured to be significantly smaller than the
same-spin~\cite{lagoudakis_stimulated_2002}. The third scenario (‘cross-loss’)
requires the condensate loss rate to increase when opposite spins are present.
If polaritons are directly excited by light, observations have suggested
biexciton formation can produce such enhanced
losses~\cite{paraiso_multistability_2010}.  However, this is only significant
when the relative spectral detuning of cavity and exciton is small
($<\SI{2}{\milli\electronvolt}$), which is far from the case here. We observe
the spontaneous buildup of circular polarization throughout the entire $+2$ to
\SI{-10}{\milli\electronvolt} detuning range. In the remaining scenario (`spin
bifurcation'), which we present in Section~\ref{sec:theory} for the first time,
the buildup of circular polarization is caused by small differences of the
energy and dissipation rates of two orthogonal linearly-polarized polariton
modes, which is present even at $k=0$.

Strain-induced splitting of the linear
components of polariton condensates has been demonstrated
previously~\cite{balili_huge_2010,klopotowski_optical_2006}. In our sample this splitting varies depending on the position on the sample. We observe a linear
polarization energy splitting of up to $\SI{100}[\sim]{\ueV}$ depending on the
position of excitation on the membranes.  Note that, all the effects reported here are also
seen on unetched samples, so strain from patterning is not crucial. However, we
observe a higher splitting at the edges of the membrane than in the middle, as
there is more stress in the structure at the edges. Due to the curvature of the cavity stopband, any 
energy splitting is accompanied by a difference of the linewidth (dissipation
rate), as explained in Appendix~\ref{app:polsplit}.  This energy splitting
between the two linear components combined with a difference in dissipation rates
causes the polarization of the condensate to change from linear
polarization to circular at a critical density (see Section~\ref{sec:theory}).

We emphasize that the stochastic circular polarization here cannot be explained in
the framework of the optical spin Hall
effect~\cite{kavokin_optical_2005,kammann_nonlinear_2012}. In our trapping
geometry the condensate is formed at the ground state with ($\bar{k}=0,\;\delta
k=\SI{0.4}{\um^{-1}}$) [FIG.~\ref{fig:1}(c)], where the transverse-electric and
transverse-magnetic (TE-TM) splitting
vanishes~\cite{shelykh_semiconductor_2004}. The trap diameter here
is 6 times smaller than the observed spin-ring patterns measured for a
nonequilibrium condensate formed at much higher in-plane wave-vectors in the
same sample~\cite{kammann_nonlinear_2012}. Moreover, the geometry or the
orientation of the trap does not affect the polarization state of the
condensate. We also see the stochastic circular polarization with a ring-shaped trap
and also in high-order spatial mode condensates~\cite{cristofolini_optical_2013,
    askitopoulos_robust_2014}. Finally, it should be noted that any theoretical
picture that assumes the buildup of circular polarization arises only because of 
the geometrical arrangement of the pump, would necessarily fail to explain the
most essential part of this work, which is the spontaneous  symmetry breaking
(stochastic behavior).

\section{Spin Bifurcation Theory (Broken Parity)\label{sec:theory}}

\begin{figure*}[htb]
    \includegraphics[width=1
    \textwidth]{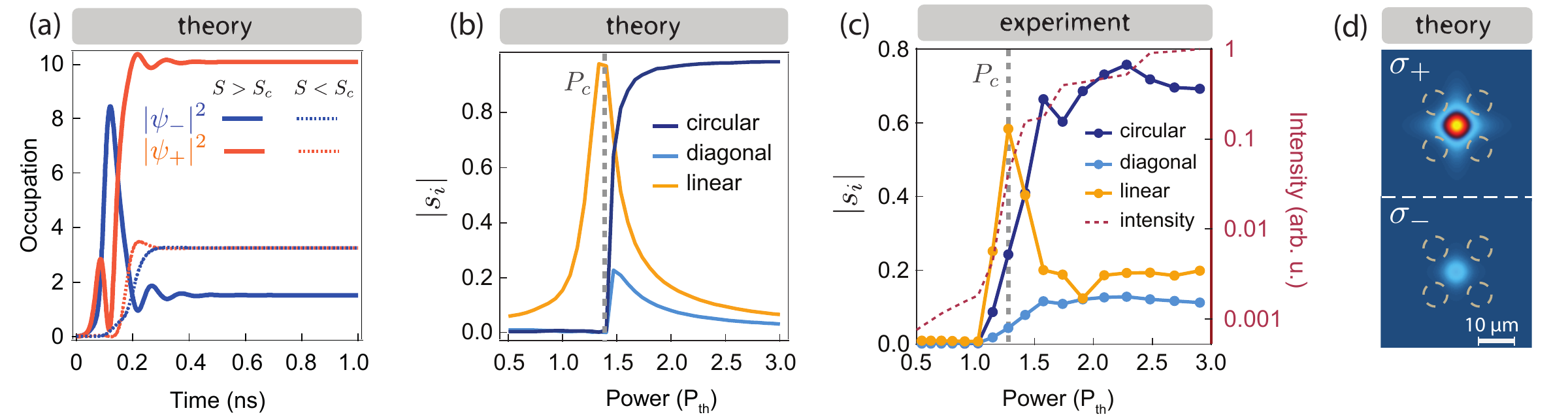} \caption{
		(a) Numerical calculations using Eq.~\ref{eq:pseudospinEq} for the case when the
        condensate occupation is below the critical occupation $S_c$ (dotted lines)
        and when above $S_c$ (solid lines). Blue lines show the occupation of the spin-down
        component and orange lines show the spin-up component. (b) Numerical calculations of the Stokes parameters at
        different powers using Eq.~\ref{eq:pseudospinEq}. (c) Measured power dependence of
        the condensate Stokes parameters. (d) 2D
        simulations, with asymmetric initial conditions [$s_z(t=0)<0.01$] show
        density of the $\psi_\pm$ components of the wave function. Pump spots
        are marked by dashed lines (see Appendix~\ref{app:simparam} for
        parameters).
    } \label{fig:theoryPD}
\end{figure*}

Our theory is a development of the theory of polariton weak lasing in two
coupled condensation centers~\cite{aleiner_radiative_2012} now for the case of
the spin degree of freedom. Here, we have right and left circular polarizations
instead of two separated condensates, and we also allow for the gain-saturation
nonlinearity in the system.

The order parameter for an exciton-polariton condensate is a two-component
complex vector $\Psi=[\psi_{+1},\psi_{-1}]^\mathrm{T}$, where $\psi_{+1}$ and
$\psi_{-1}$ are the spin-up and spin-down wave functions. The components of the
order parameter define the measurable condensate pseudospin
$\mathbf{S}=(1/2)(\Psi^\dag\cdot\bm{\sigma}\cdot\Psi)$, and the normalized spin vector
$\mathbf{\hat{s}}=\mathbf{S}/S$, where $\sigma_{x,y,z}$ are the Pauli matrices.
The components of this vector contain information about the intensities and
relative phases of the emitted light. The order parameter evolves according to the driven
dissipative equation
\begin{align}
    \label{eq:main}
    \begin{split}
        \mi \frac{d\Psi}{dt}=&-\ihalf
        g(S)\Psi-\frac{\mi}{2}(\gamma-\mi\ve)\sigma_x\Psi
        \\&+\half\left[(\alpha_1+\alpha_2)S+(\alpha_1-\alpha_2)S_z\sigma_z\right]
    \Psi,
\end{split}
\end{align}
or in components:
\begin{subequations}
\begin{alignat}{2}
    \begin{split}
        \dot{\psi}_{+1}=&-\half g(S)\psi_{+1}
        -\half(\gamma-\mi\ve)\psi_{-1}
        \\&-\ihalf
        (\alpha_1\vert\psi_{+1}\vert^2+\alpha_2\vert\psi_{-1}\vert^2)\psi_{+1},
    \end{split}\\
    \begin{split}
         \dot{\psi}_{-1}=&-\half g(S)\psi_{-1}
         -\half(\gamma-\mi\ve)\psi_{+1}
         \\&-\ihalf
         (\alpha_1\vert\psi_{-1}\vert^2+\alpha_2\vert\psi_{+1}\vert^2)\psi_{-1}.
    \end{split}
\end{alignat}
\label{eq:maincomp}
\end{subequations}
Here, $g(S)=\Gamma-W+\eta S$ is the pumping-dissipation balance, $\Gamma$ is the
(average) dissipation rate, $W$ is the incoherent in-scattering (or `harvest'
rate), and $\eta$ captures the gain-saturation term with $S = (\vert
\psi_{+1}\vert^2+ \vert \psi_{-1}\vert^2)/2$~\cite{keeling_spontaneous_2008}.
Here, this gain saturation depends on the total occupation of the
condensate (treated more generally in Appendix~\ref{app:varcross}). It is
assumed now that $X$ (horizontal) and $Y$ (vertical) linearly-polarized single-polariton
states have different energies and dissipation rates. The energy of the
$X$-polarized state is shifted by $-\ve/2$, and the energy of the $Y$-polarized
state by $+\ve/2$. The dissipation rate from the $X$-polarized state is $\Gamma+\gamma$,
while the dissipation rate from the $Y$-polarized state is $\Gamma-\gamma$ (see
also Appendix~\ref{app:polsplit}).  Finally, $\alpha_1$ is the repulsive interaction
constant for polaritons with the same spin, and $\alpha_2$ is the interaction
constant for polaritons with opposite spins.

From Eq.~\ref{eq:main} we obtain for the components of the pseudospin vector
($\alpha=\alpha_1-\alpha_2$):
\begin{subequations}\label{eq:pseudospinEq}
\begin{alignat}{3}
    \begin{split}
        \dot{S_x}=-g(S)S_x-\gamma S-\alpha S_z S_y,
    \end{split}\\
    \begin{split}
        \dot{S_y}=-g(S)S_y+\ve S_z+\alpha S_z S_x,
    \end{split}\\
    \begin{split}
        \dot{S_z}=-g(S)S_z-\ve S_y,
    \end{split}
\end{alignat}
\end{subequations}
and the related equation for the total spin $\dot{S}=-g(S)S-\gamma S_x$. There
are two sets of solutions, which we call here the paramagnetic and ferromagnetic
solutions.

\paragraph{Paramagnetic solutions.} These give simple condensation into either $X$ or $Y$
linearly-polarized states. The $Y$ state possesses the longest lifetime, and the condensation
threshold is reached for this state first at $W_1=\Gamma-\gamma$. There is no parity breaking for
this condensate: $S_y=S_z=0$, $S_x=-S$ with $S=(W-W_1)/\eta$, so that the occupations of $+1$ and
$-1$ components are equal. However, this condensate solution becomes unstable for $W>W_2$.
The values of the critical occupation $S_c$ and the critical pumping rate are
\begin{equation}
S_c=\frac{\gamma^2+\ve^2}{\alpha\ve},\qquad W_2=W_1+\eta S_c.
\end{equation}
Note that this instability is present also for equal dissipation rates, i.e. when $\gamma=0$. In
this case, the system (Eq.~\ref{eq:pseudospinEq}a-c) describes the self-induced Larmor precession of
the pseudospin vector. Incorporating energy relaxation (e.g., using small negative $\gamma$) then 
leads to the formation of the $X$-polarized condensate---an intuitively expected result.

\paragraph{Ferromagnetic solutions.} The key ingredient of our theory is the
presence of the $\gamma>0$ parameter describing the variation of dissipation
rates. This parameter allows the formation of the `weak lasing' regime
\cite{aleiner_radiative_2012}, which is characterized by two important features:
(i) the $X$-polarized condensate is also unstable, and (ii) when the
$Y$-polarized condensate loses stability at the critical occupation $S_c$, it
continuously transforms into one of the two ferromagnetic states. While
Eqs.~\eqref{eq:maincomp} are parity symmetric, i.e., they are not affected by
the interchange of left and right circular polarization, the new solutions are
characterized by broken parity symmetry and by spontaneous formation of either
left or right elliptical polarization. These solutions are
\begin{subequations}\label{eq:Sz_pos}
    \begin{equation}
        S_x=-\frac{g(S)}{\gamma}S, \qquad S_y=-\frac{g(S)}{\ve}S_z,
    \end{equation}
    \begin{equation}
        S_z=\pm\frac{\ve}{\gamma}\sqrt{\frac{\gamma^2-g(S)^2}{\ve^2+g(S)^2}}\,S, \qquad
        S=\frac{\gamma[\ve^2+g(S)^2]}{\alpha\ve g(S)},
    \end{equation}
\end{subequations}
where the positive root of the second equation in Eq.~(\ref{eq:Sz_pos}b) should be taken. We note
that while the sign of $S_x$ is always negative, $S_y$ and $S_z$ have opposite signs for the two
solutions. This means that the left-circular component is accompanied by a diagonal component, and
the right-circular by anti-diagonal. Moreover, if these components change for some reason, they
mirror each other as long as the total condensate occupation stays fixed. We label these two
solutions as the $s_\downarrow$ and $s_\uparrow$ spin states.

The spin-independent model for the gain saturation used in this Section is sufficient to
describe the experimentally observed features. However, the spin relaxation in the
reservoir can be slow~\cite{li_incoherent_2015}, and in this case the model should be modified to
allow the saturation terms to depend on the individual occupations of the left- and the right-circular
polarization components rather than the total occupation only. The parity breaking is
still present after this modification; however, the stability of solutions becomes more complex. The
ferromagnetic solutions can now become unstable and transform into periodic cycles~\cite{rayanov_frequency_2015},
the dynamics of the pseudospin can become irregular, and this can
also result in the formation of the stable $X$-polarized condensate at high pumping powers. See 
Appendix~\ref{app:varcross} for more details.

Numerical calculations for the occupation of the two circular components of the wavefunction when
$S<S_c$ (dotted lines) and when $S>S_c$ are shown in FIG.~\ref{fig:theoryPD}(a). Here, the
condensate is initialized with a small asymmetry in spin-up and spin down occupations (<1\%). Below
the critical occupation $S_c$, the condensate is linearly polarized, but when the occupation is
increased above the threshold $S_c$, the condensate adopts one of two elliptically polarized
configurations depending on the initial conditions. In the experiment, the stochastic behavior is due to random spin fluctuations at the onset of the condensation. In theory, we can reproduce it by randomly setting the initial conditions.  Numerical calculations of the condensate polarization versus excitation power
are shown in FIG.~\ref{fig:theoryPD}(b). Directly at the condensation threshold the condensate is
linearly polarized, but once it reaches the critical occupation ($P_c=1.3 P_{\mathrm{th}}$, marked
by a dashed grey line), the linear component is quenched and circular polarization builds up. This
behavior reproduces the experimental data, as shown in FIG.~\ref{fig:theoryPD}(c). We observe an
initial buildup and subsequent quenching of linear polarization with the continuing increase of 
circular polarization at $P_c=1.25P_{th}$ (marked by a dashed grey line, with total intensity marked by
a dotted red line). Once circular polarization is achieved, the orientation of the condensate
circular polarization becomes stochastic under linearly polarized pumping.

We can extend Eq.~(\ref{eq:main}) and account for 2D real-space degrees of freedom by using complex
Ginzburg-Landau-type equations~\cite{wouters_excitations_2007,
keeling_spontaneous_2008,dreismann_coupled_2014}, which in addition to the pump and decay also
incorporate a repulsive potential due to the excitons in the pump spots and an energy
relaxation~\cite{wouters_energy_2010} for polaritons in the trap:
\begin{multline}
 \mi\frac{d\Psi}{dt}=-\frac{\mi}{2} \left[ g(S)+ \gamma\sigma_x \right]\Psi \\
 +(1-\mi\Lambda)\Big\{
 \frac{1}{2}\left[(\alpha_1+\alpha_2)S+(\alpha_1-\alpha_2)S_z\sigma_z\right]\Psi \\
 - \frac{1}{2}\ve\sigma_x\Psi-\frac{\nabla^2}{2m^*}\Psi+V_p\Psi \Big\},
\end{multline}
where $m^*$ is the effective mass of the polaritons. The harvest rate is given by $W=rP/\Gamma_R$,
where $P$ is the spin-independent spatial profile of the excitation, $\Gamma_R$ is the decay rate
of the exciton reservoir, and $r$ is the incoming rate of polaritons into the condensate. The gain
saturation is given by $\eta=r^2 P/\Gamma_R^2$. The repulsive potential due to the interaction of
polaritons with the exciton reservoir is given by $V_p=\half g_r N + \half g_P P$, where $g_r$, and
$g_P$ are the interaction constants of polaritons with the exciton reservoir and the pump spot
respectively, and $N=g(S)/r$ is the density of the exciton reservoir. Here, $\Lambda\ll 1$ is a
phenomenological constant that gives the energy relaxation.  The density profile of the two
circular components of the wave function in the steady state [FIG.~\ref{fig:theoryPD}(d)], for the case of
a trapped condensate in the middle of the 4 pump spots, exhibits a circular polarization degree of
$\vert s_z \vert=0.69$. Note that with $\gamma=0$ and only polarization splitting (including TE-TM splitting), our 2D simulations do \emph{not} show bistable condensation.

It is important to note the differences between the ferromagnetic states we
discuss here and the magnetization transition in equilibrium cold atom
systems~\cite{sadler_spontaneous_2006,stamper-kurn_spinor_2013}.  First, the
parity breaking bifurcation described above does not reduce the energy of the
system (unlike for atoms). In fact the energy of elliptically polarized states is {\it higher} than that of linearly polarized states. Second, the in-plane components of the spin do not vanish
completely. Third, we have a magnetized condensate with only two possible
orientations,  whereas in atomic systems ferromagnetic domains with
continuous variable orientation are observed.

If the Hamiltonian and the initial state of a system are symmetric under the
exchange of spin-up and spin-down components, but the final state is not, the parity
symmetry is spontaneously broken. This is indeed the case here: We excite an
equal population of spin-up and spin-down polaritons, which spontaneously
condense, but form highly circularly polarized macroscopic states in the
absence of any external magnetic field.

\section{Resonant excitation\label{sec:res}}

The GaAs substrate commonly used in the fabrication of GaAs microcavities is
opaque at the emission wavelength ($\SI{\sim800}{\nano\meter}$) of the cavity
polaritons. As a result, the back side of the cavity is resonantly inaccessible.
Resonant excitation of the cavity from the front side has complications with
backscatter from the laser, especially in high finesse cavities and at normal
incidence, where the condensate emission mode is located. To circumvent this
problem we chemically etch the substrate to form membranes of \SI{8.8}{\um}
thickness and \SI{300}{\um} diameter [see FIG.~\ref{fig:1}(a)]. For resonant excitation, we use a narrow-linewidth ($<\SI{2}{\GHz}$) CW laser, which is amplitude
modulated with a second AOM. We call this resonant laser the `gate'. We use two photomultipliers and a fast
oscilloscope to time resolve the left- ($\sigma_-$) and right-circular
($\sigma_+$) polarization intensity  of the condensate emission.  The resonant
excitation laser, the nonresonant pump laser, the cameras and the oscilloscope
are all synchronized, which allows us to vary the delay time and
amplitudes of each laser pulse on demand.

\subsection{Resonant initialization of spin states}

\begin{figure}[htp]
	\includegraphics{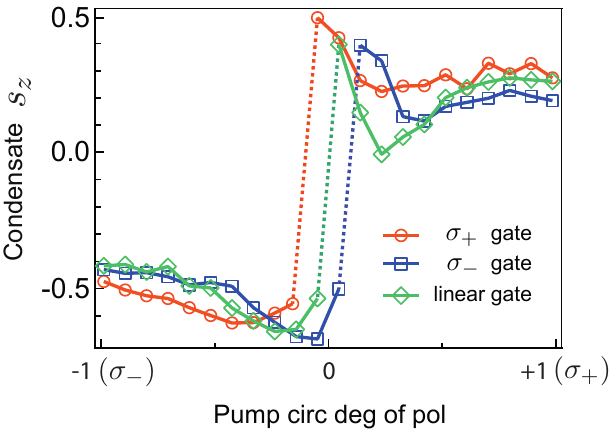} \caption{
   	    Measured
        circular polarization of condensate vs pump
        ellipticity, for three different gate polarizations. Dashed lines
        mark the region where the condensate initializes stochastically in 
        spin-up or spin-down states in each realization.
    \label{fig:initialization} } 
\end{figure}

We can control the polarization of the trapped condensate with the gate
laser. In this case, we additionally \emph{resonantly} excite the condensate (which is generated by
the four nonresonant pump spots) with a second laser from the backside of the microcavity membrane. This gate can be
linearly, left-, or right-circularly polarized. FIG.~\ref{fig:initialization} shows the
condensate circular degree of polarization versus that of the pump. In the green curve which shows the behavior when the gate is linearly polarized, we reproduce
the same result as that in FIG~\ref{fig:pulsetrain}(d). However, when a right (or left) circularly
polarized gate is applied, the curve shifts to the left (or right). In other
words, with a linearly polarized pump, the condensate initializes in a right (or
left) circularly polarized state. The imbalance caused by the resonant gate
cancels out with an opposite circularly polarized pump with a circular
polarization degree of $\vert s_{z,\text{pump}} \vert\sim 0.1$.

\subsection{Elliptically-polarized pump: coherent driving}

\begin{figure*}
	\centering 
	\includegraphics[width=1
	\textwidth]{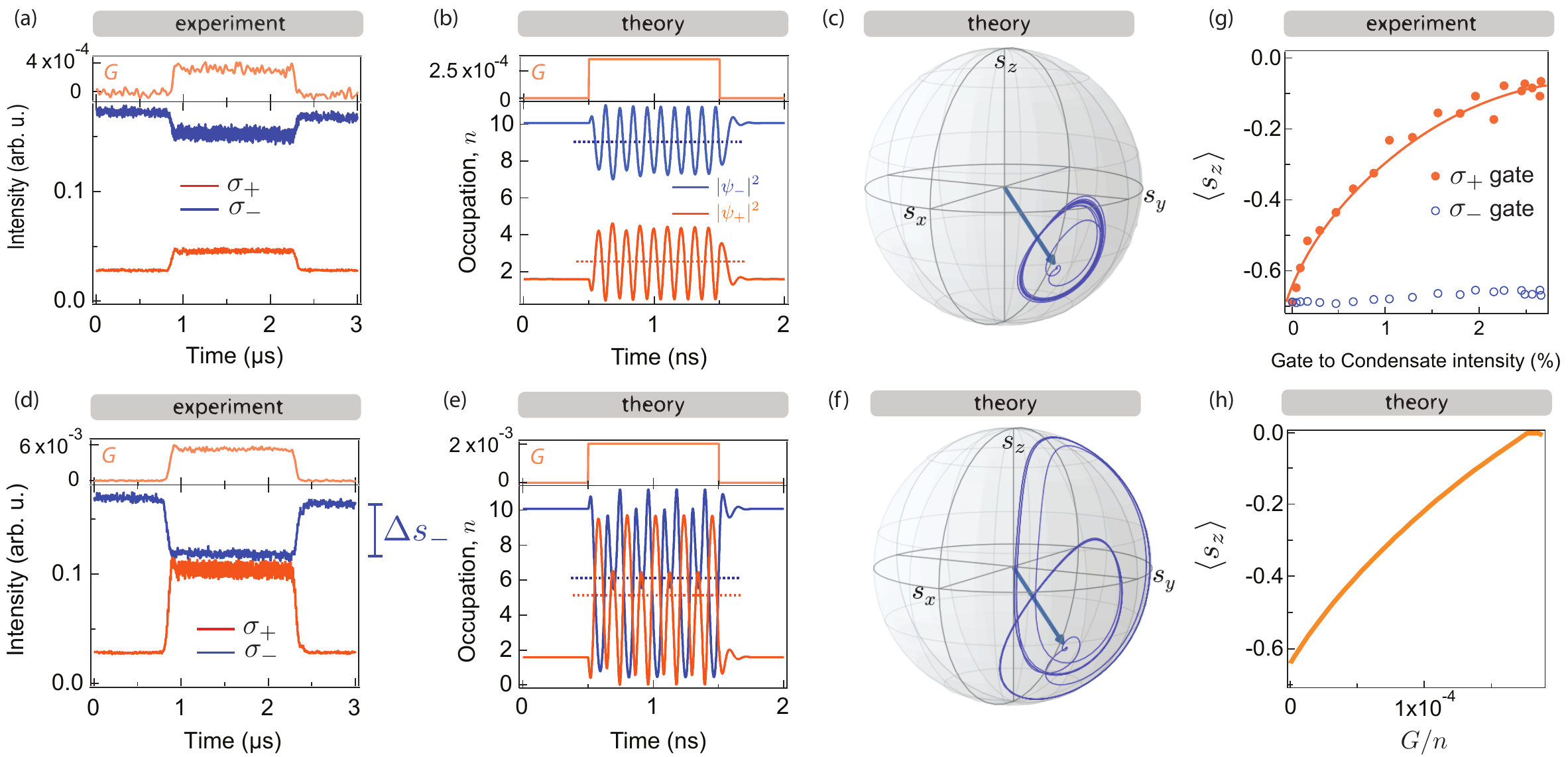} \caption{
        (a) Measured intensity of circular polarization components
        ($\sigma_\pm$) for adding gate pulse $G$ (intensity shown on above inset). The
        condensate initializes in the $s_\downarrow$ state, due to the asymmetric pumping
        condition $s_{z,\text{pump}}=-0.1$. A gate pulse with an opposite
        circular polarization ($\sigma_+$) is applied. (b) Numerical calculations for
        (a), showing that during the gate pulse $G=A^2$ the spin-up and spin-down
        densities oscillate with a $\pi$ phase shift. (c) Time evolution of
        the pseudospin, showing how the pseudospin orbits around the spin-down
        stationary state (marked by the blue arrow) when the gate pulse is applied.
        (d) Same as in (a) but with a fivefold increase in the intensity of the
        gate. The time-averaged condensate emission is almost circularly unpolarized when the pulse is applied
        ($\langle s_z \rangle=0$). (e) Numerical calculations for (d), showing almost
        equal amplitude oscillation of the spin-up and spin-down densities. (f)
        Trajectory of pseudospin, showing that with a strong gate pulse large
        orbits commence around the two spin-up and spin-down stationary states.
        (g) Average circular polarization when the gate is on, at different gate
        intensities. When an opposite polarization gate is applied,
        $\langle s_z \rangle$ converges to zero, whilst it remains unchanged for
        the same-spin gate. The orange line is a guide to the eye. (h) Numerical calculations of $\langle s_z \rangle$ vs
        normalized pump intensity ($n$ is the condensate occupation).
    } 
   \label{fig:res}
\end{figure*}

In a first experiment, we initialize a condensate in the spin-down state
($s_\downarrow$) by making the pump laser slightly left-circularly polarized
(`asymmetric pumping'). Once the condensate is created, we excite the sample
resonantly from the back side with an oppositely circularly polarized
($\sigma_+$) gate.  FIG.~\ref{fig:res}(a) shows the intensity of $\sigma_+$ and
$\sigma_-$ components of the condensate emission during this gating. We see a
reduction in intensity for the component which is opposite to the gate laser
polarization and an increase for the same spin polarization component.  To
account for this in the theory, we add a new term to Eq.~\ref{eq:maincomp},
corresponding to the resonant laser:
\begin{subequations}
\begin{alignat}{2}
    \begin{split}
        \dot{\psi}_{+1}=&-\half g_{+1}(S)\psi_{+1}
        -\half(\gamma-\mi\ve)\psi_{-1}
        \\&-\ihalf
        (\alpha_1\vert\psi_{+1}\vert^2+\alpha_2\vert\psi_{-1}\vert^2)\psi_{+1}\\
        &-\mi A\,\Pi(t,t_0,\delta t)e^{-\mi\omega_g t},
    \end{split}\\
    \begin{split}
        \dot{\psi}_{-1}=&-\half g_{-1}(S)\psi_{-1}
         -\half(\gamma-\mi\ve)\psi_{+1}
         \\&-\ihalf
         (\alpha_1\vert\psi_{-1}\vert^2+\alpha_2\vert\psi_{+1}\vert^2)\psi_{-1},
    \end{split}
\end{alignat}
\label{eq:mainCompRes}\end{subequations}
where $A\,\Pi(t,t_0,\delta t)=AH(t-t_0)H(t_0+\delta t-t)$ uses the Heaviside
step function $H$ to give a square pulse with amplitude $A$ which starts at time
$t_0$ and lasts for $\delta t$, and $\omega_g$ is the excitation frequency. To
account for the elliptically-polarized pumping we modify $g(S)$ to $g_{\pm
    1}(S)=\Gamma-W_{\pm 1}+\eta S$. Numerical calculations
[FIG.~\ref{fig:res}(b)] show the case $W_{-1}=1.09\;W_{+1}$, $A=5\times 10^{-4}
n$, where $n$ is the occupation of the condensate, and $\omega_g=-0.3 \ve$. The
dotted line marks the time average of the oscillations. The condensate
pseudospin vector precesses around the stationary state pseudospin vector at a
frequency of $\omega_L/2\pi\sim\SI{10}{\GHz}$ in a limit cycle [see
FIG.~\ref{fig:res}(c)]. In the case where the condensate is highly
spin-polarized and the gate is at resonance ($\omega_g=0$), the oscillation
frequency is equal to the self-induced Larmor precession frequency
$\omega_L=\gamma\varepsilon/g$ (Appendix~\ref{app:kramers}). This sets the
fastest possible spin dynamics in the system, and does not depend on the gate
intensity. Here, the condensate pseudospin oscillates faster than our detection
time resolution and consequently we only see the average effect in
FIG.~\ref{fig:res}(a). The coherent driving reported here
strongly depends on the detuning of the gate laser frequency relative to that of the
condensate. The resonance width at which we can drive the condensate is determined
to be 10-\SI{20}{\ueV}, as explained in Appendix~\ref{app:reswidth}.

\begin{figure*}
	\centering 
    \includegraphics[width=1\textwidth]{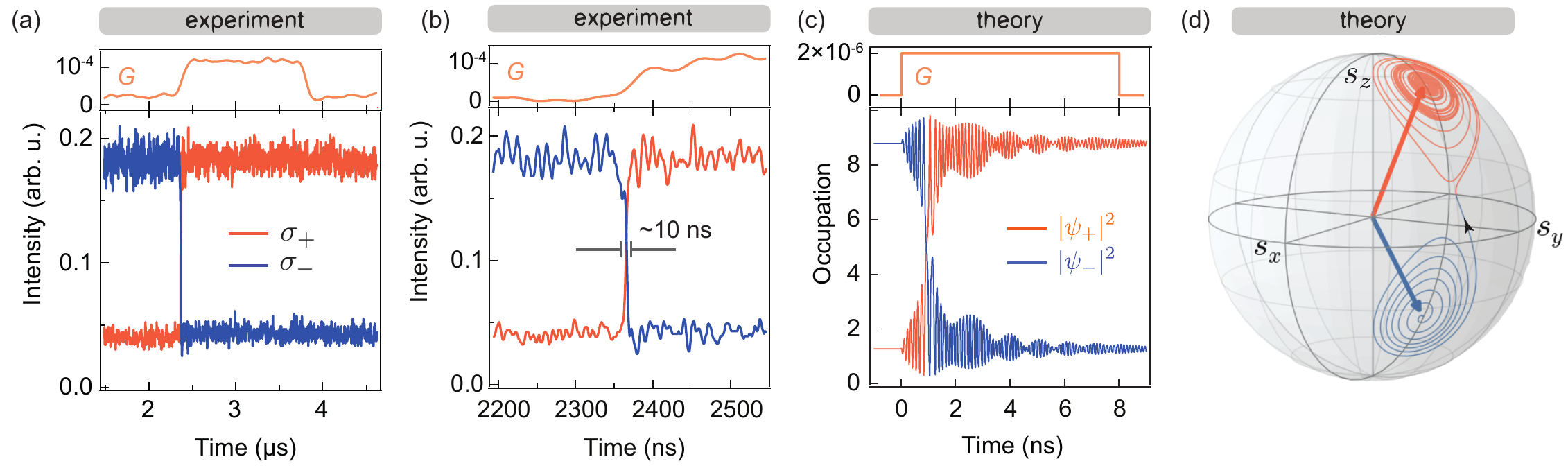} \caption{
      (a) Intensity of the spin-up and -down states as a function of time. Resonant
      pulse $G$, with opposite circular polarization
      ($\sigma_+$) switches the circular polarization of the
      condensate. (b) Despite the slow rise time of our AOM ($\SI{\sim
          100}{\nano\second})$, the condensate switches to the opposite spin
      state in less than $\SI{10}{\nano\second}$ (the limit of our detection),
      as soon as the intensity crosses the spin-switching threshold. (c) Numerical
      calculations for (a), showing that the spin switches with a gate pulse
      with intensity 6 orders of magnitude smaller than that of the condensate.
      The evolution of the pseudospin is shown in (d). The two attractors,
      marked by the blue and orange arrows, are the two stationary
      $s_{\uparrow\downarrow}$ spin-polarized states.
   } \label{fig:resSwitch} 
\end{figure*}

In the case of FIG.~\ref{fig:res}(d) the intensity of the gate laser is
increased by five times with respect to that in (a). Under these conditions, the
time average of the circular component of the condensate emission becomes almost
unpolarized. If the amplitude of the resonant excitation is large enough, the
condensate pseudospin can cross over to the second attractor (the spin-up state)
and form a large trajectory which encompasses both spin-up and spin down
stationary states with a characteristic period
doubling~\cite{alexeeva_actively_2014} [FIG.~\ref{fig:res}(e,f)]. The dotted
lines in (e) mark the time average of the oscillations. FIG.~\ref{fig:res}(g)
shows the power dependence of the average circular polarization as a function of
the intensity of the gate. As the oppositely polarized gate intensity is
increased, we see the condensate average circular polarization converges to
zero. Here, we have the gate resonant with the condensate. However, the coherent
driving strongly depends on the detuning of the gate laser frequency to that of
the condensate. The resonance width where we can drive the condensate is
determined to be 10-\SI{20}{\ueV}, as explained in Appendix~\ref{app:reswidth}.
That theory and experiment agree regarding the observed average polarization
over the whole range of gate powers, and also the observation of a resonance
(see Appendix~\ref{app:reswidth}) strongly suggests that our description in
terms of coherent driving is valid.

In summary, the fixed points of broken parity symmetry
($s_\downarrow,s_\uparrow$) become unstable due to small perturbations and
convert to limiting circles around stationary states. The precession is linear
for small gate amplitudes but it becomes nonlinear for large amplitudes. This is
manifested by oscillation of the condensate parameters, occupation and
polarisation with time. In this linear regime, the gate pulse appears to act as if it induces an effective magnetic field, around which the condensate polarisation precesses. It is worth noting that this system can exhibit chaotic behavior with increased amplitude of the pulse. Increasing the amplitude first
results in period doubling, and eventually leads to chaos (Feigenbaum
scenario)~\cite{alexeeva_actively_2014}. The ability to coherently control the
spin of the condensate suggests its utility for computational operations. 

\subsection{Linearly-polarized pump: spin switching}

We now explain the case when the nonresonant pump is linearly polarized. In this
case there is no external `force' to drive the condensate to a specific spin
state. As a result, once the circular component of the condensate pseudospin
crosses zero, it falls into the opposite spin state's attractor.
FIG.~\ref{fig:resSwitch}(a) shows a realization which lasts for
$\SI{4}{\micro\second}$ with the condensate first initialized in the spin down
state ($s_\downarrow$). We resonantly excite the condensate with an
oppositely polarized gate ($\sigma_+$). This causes the switching of the polarization
of the condensate to the same polarization as that of the gate. The gate, which
is $\SI{1.5}{\micro\second}$ in duration, is then turned off but the condensate
remains in the switched polarization because the pump is linearly polarized and
the symmetry is not explicitly broken. This shows that we are capable of
manipulating the polarization of the condensate on demand, while also
reinforcing our observation that the condensate picks a polarization
spontaneously when it is formed. It should be noted that the state of the
condensate does not change when the circular polarization of the gate is the
\emph{same} as that of the condensate.  

If the symmetry breaking was somehow set by the parameters of the experiment,
as in the previous example where the pump was slightly elliptical, the
condensate would switch back to its original state after the gate beam was
turned off. Instead, a very small gate field in the cavity can initiate
coherent switching. The upper panel in FIG.~\ref{fig:resSwitch}(a) shows the
transmitted gate intensity when the pump is blocked. Indeed, the gate power
before the cavity is \SI{600}{\micro\watt}, which is 50 times weaker than the
pump power. However, only a small fraction of this resonant beam couples to the
cavity in our setup.  Therefore, a suitable comparison of the intensity of the
gate to that of the condensate is to compare the transmitted intensity of the
gate to the intensity of the condensate in the same direction. We find that the
gate intensity is more than 60 times weaker than the condensate intensity, when
the gate laser frequency is tuned to the bottom of the polariton dispersion
(accounting for the blueshift of the condensate).

By estimating the occupation number of the condensate (see
Appendix~\ref{sec:particlenumber}) we find that only 13 polaritons are enough to
reverse the spin state of the condensate. The condensate switching time is less
than \SI{10}{ns} (our detection limit), and is 10 times faster than the
switching time of the AOM [FIG.~\ref{fig:resSwitch}(b)]. This important fact
shows that the condensate does not follow the optical gate adiabatically.
Simulations of the spin-switching with minimum resonant gate intensity are shown
in FIG.~\ref{fig:resSwitch}(c).  The polarization of the gate is opposite to
that of the condensate and we have assumed symmetric pumping rates for the
spin-up and spin down condensates i.e.  $W_{+1}=W_{-1}$. The condensate
polarization reverses within $\SI{\sim200}{\ps}$ once the gate is applied and
remains in that state after the pulse is turned off during continued spin
evolution [FIG.~\ref{fig:resSwitch}(d)]. Switching requires a minimum gate intensity, to twist the condensate pseudospin onto the equator in the Poincaré sphere. This sets a threshold for the gate power.  By measuring the gate flux, and using
the switching time of $\SI{10}{\ns}$, an \emph{upper} bound of the minimum energy for
switching the state of the condensate is found to be $\SI{1E-15}[\sim]{J}$,
comparable to state-of-the-art optoelectronic switches with similar
speeds~\cite{nozaki_sub-femtojoule_2010}. We emphasise, however, that the
theoretical limit for the minimum switching energy is 50 times smaller
because the spin dynamics of the system ($\sim\SI{200}{\ps}$) is 50 times faster
than our detection limit.  This polaritonic system is then an extremely low-power switch.

Multistability has indeed been demonstrated before in resonantly pumped
condensates ~\cite{paraiso_multistability_2010,cerna_ultrafast_2013}. However,
there are several major differences between our system and that of the Deveaud
group. In Deveaud's experiments the physical process causing multistability is the
nonlinear nonradiative losses in the polariton gas due to the formation of
biexcitons. These losses are only significant when the polariton gas energy is
close to the biexciton energy
($<\SI{2}{\milli\electronvolt}$)~\cite{wouters_influence_2013}. In contrast here, the
bistability that we observe is present $>\SI{10}{\milli\electronvolt}$ below the
biexciton energy. There is no phase transition in Paraïso et al. work~\cite{paraiso_multistability_2010} on
multistability and symmetry is \emph{not} spontaneously broken. In contrast, we observe spin symmetry breaking while the power thresholds of
the left- and right-circularly polarized states are the same, allowing us to
observe spontaneous magnetization. In the Deveaud multistable system, in order
to switch the polarization one has to inject an opposite-spin polariton density
equal to the density difference of spin-up and spin-down polaritons.  In contrast, in our experiments, the
condensate switches with a gate intensity 60 times weaker than the condensate.
Theoretically, the gate intensity can be much weaker and the reason that we do
not see switching at even weaker powers experimentally is due to spin
fluctuations in the condensate, as discussed below. Finally, we note that other
microstructured systems such as microdisk lasers can also show bistability.
The whispering gallery modes in microdisk lasers can have clockwise or
counterclockwise propagating lasing modes~\cite{hill_fast_2004}.  When the
coupling between the two modes is small, a cross-gain saturation causes one of
the modes to dominate. The system can therefore operate in ``flip-flop'' mode
and the two modes can be excited and switched with weak optical
pulses~\cite{liu_ultra-small_2010}.

\subsection{Thermal noise: spin flipping\label{sec:thermnoise}}

\begin{figure}[!htbp]
	\centering 
	\includegraphics[width=0.48
	\textwidth]{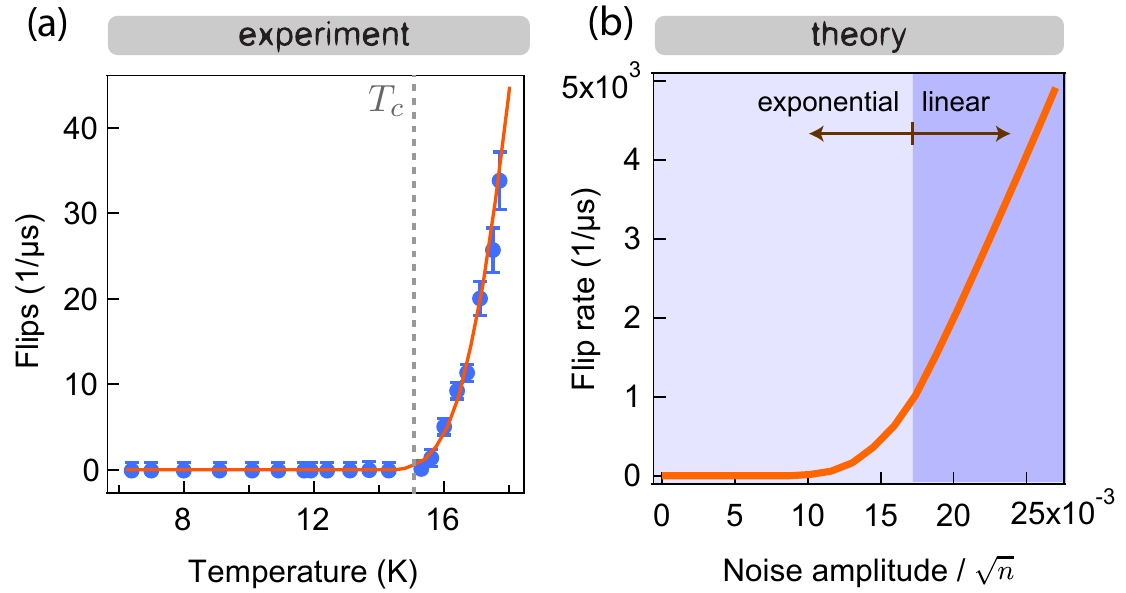} \caption{
        (a) Measured spontaneous spin flip rate at different temperatures of
        the sample. Above the temperature
        $T_c\simeq\SI{15}{\K}$, the flip rate increases exponentially. The
        solid line shows the theoretical fit. (b) Calculated dependence of flip
        rate with normalized noise amplitude. The flip rate increases
        exponentially as the noise amplitude increases (light region), and eventually becomes
        linear (dark region). Comparison with (a) shows that the measured flip rates remain
        below this linear regime.
    } \label{fig:temp} 
\end{figure}

Above the spin bifurcation occupation threshold ($S_c$), any small perturbation
induces a `restoring force', which drives the condensate toward the attractor
once the perturbation is stopped. This restoring force keeps the condensate
around the attractor for small spin fluctuations. However, for sufficiently large
perturbations, the condensate can flip to the other attractor with the
opposite circular polarization. In the experiment, for sample temperatures
near \SI{5}{\K} the observed spin-polarized states remain stable for many seconds (longer than the stability of our experiment can be maintained).
However, increasing the temperature above $T_c\simeq\SI{15}{\K}\pm1$ induces spin flips in the condensate during a measurement time window of $t_m=\SI{1.5}{\ms}$. The measured rate of the spin flipping beyond $T_c$ is a nonlinear function of the sample
temperature as shown in FIG.~\ref{fig:temp}(a). It is important to note that
we do not have a real threshold for the occurrence of spin flipping here. In
fact, the spin-flip rate increases exponentially with temperature right from
$T=0$, until it becomes significant in the finite-time measurement window. We can account for this phenomenon by adding a thermally induced noise to our theory. This thermal noise $f(t)$, which is similar to the Johnson noise, is included in Eq.~(\ref{eq:maincomp}) to
give Langevin-type equations:
\begin{subequations}\label{eq:noise}
\begin{alignat}{2}
    \begin{split}
        \dot{\psi}_{+1}=&-\half g(S)\psi_{+1}
        -\half(\gamma-\mi\ve)\psi_{-1}
        \\&-\ihalf
        (\alpha_1\vert\psi_{+1}\vert^2+\alpha_2\vert\psi_{-1}\vert^2)\psi_{+1}+f_{+1}(t),
    \end{split}\\
    \begin{split}
         \dot{\psi}_{-1}=&-\half g(S)\psi_{-1}
         -\frac{1}{2}(\gamma-\mi\ve)\psi_{+1}
         \\&-\ihalf
         (\alpha_1\vert\psi_{-1}\vert^2+\alpha_2\vert\psi_{+1}\vert^2)\psi_{-1}+f_{-1}(t),
    \end{split}
\end{alignat}
\end{subequations}
where $f_{\sigma}(t)$ with $\sigma=\pm1$ is a realization of Gaussian
random processes with zero mean $\langle f_\sigma(t)\rangle=0$ and $\delta$-like
two-point correlation function
\begin{equation}\label{eq:noise2}
 \langle f_\sigma(t) f_{\sigma'}(t')\rangle=0, \quad
 \langle f_\sigma(t) f^*_{\sigma'}(t')\rangle=2 D\delta_{\sigma,\sigma'}\delta(t-t').
\end{equation}
At finite temperature $T$ the intensity of the noise can be written
approximately as $D=\half (W+aT)$, with the $W$-contribution being a shot noise from the reservoir~\cite{aleiner_radiative_2012}
and the thermal part $aT$ defined by spin-flip polariton-phonon scattering.

The flip rate vs noise amplitude using stochastic simulations
[FIG.~\ref{fig:temp}(b)] reveals an Arrhenius-like increase at a critical
threshold followed by a crossover to a linear regime.  This model gives an
excellent account of the dynamics [FIG~\ref{fig:temp}(a)]. For temperatures
beyond \SI{18}{\K} we reach the time resolution of our detection. As a result,
we cannot completely span the crossover to the linear spin-flip regime in our experiment.
A fit of the simulation results to the experimental data in
FIG.~\ref{fig:temp}(a) gives $a=\SI{.17}{\per\ps\per\K}$,  which sets the
dependence of spin flip rate on $D(T)$. We can then explore how this noise
perturbs the spin system given by equations~(\ref{eq:noise}). 

Inclusion of noise produces a spin flip rate that can overcome the effective
spin potential barrier (see Appendix~\ref{app:kramers}). For the case in which the
circular degree of polarization is high, the spin-flip process can be considered
as a one-dimensional Kramers transition. The spin-flip rate $R_K$ can then be
estimated as
\begin{equation}\label{eq:kramers_rate}
 R_K=\frac{\sqrt{\varepsilon g(S)}}{2\pi}
 \exp\left\{ -\frac{g(S) n}{4 D} \ln\left(\frac{\ve}{g(S)}\right) \right\}.
\end{equation}
We note that for $\varepsilon=\SI{30}{\ueV}$ and the condensate occupation
$n=800$, the zero-temperature shot-noise spin-flip rate (set by $W$) is
negligible for our observation timescales (\SI{10}{\s}). The critical
temperature $T_c(t_m)$ given by $R_K(T_c) = 1/t_m$ depends dramatically on the
measurement window time $t_m$. While the phonon-polariton interaction $a$ in our
system gives $T_c(\SI{1.5}{\milli\s})=\SI{15}{\K}$, at lower temperatures the
condensate spin lifetime rapidly exceeds the stability time of our experimental
apparatus (many seconds). Modifying the phonon-polariton interaction $a$ thus
has an enormous effect on the spin stability. Finally, we note that in a similar fashion to thermal noise, an overlapping reservoir can also induce spin noise in the condensate. The spin noise causes condensate spin flips, which result in the reduction of the time-averaged circular polarization. Theoretically, this can be studied by introducing a noise term similar to Eq.~\ref{eq:noise2}, but instead of depending on temperature the noise intensity depends on the overlapping reservoir density~\cite{read_stochastic_2009}.

\section{Concluding remarks\label{sec:conc}}

In summary, we showed how spin can emerge spontaneously in nonresonantly-pumped polariton
condensates. We found that for trapped condensates, in the case where the pump
is linearly polarized, parity symmetry is spontaneously broken by spin
fluctuations at the onset of condensation.  Fluctuations are amplified by
nonlinearities in the condensate formation due to the energy and lifetime
splitting of the linear polarization components, producing a spin-up or
spin-down condensate. The symmetry can be explicitly broken by applying a
slightly elliptically polarized pump, which increases the likelihood of forming
condensates with the same spin as the pump. In the case where the pump is
linearly polarized, we switched the condensate state using a 60-fold weaker
resonant gate pulse with an opposite circular polarization.  This situation
changes when the pump is elliptically polarized, where instead of switching, the
condensate pseudospin precesses around stationary states in limit cycles.
Finally, we showed how  thermal excitations can induce spin flips with a rate
that increases exponentially with sample temperature.

We demonstrated here one way to explicitly break symmetry utilizing elliptical
polarization pumping geometries. One could also break the symmetry by introducing
a magnetic field to split the energy of the spin-down and spin-up condensates.
Alternatively, the pumping symmetry could be broken with spin current injection.
These experiments thus exhibit rich physics with potential applications in
sensing.

The observation of spontaneous discretized spin-polarized states also has interesting
consequences in the physics of condensate lattices. The possibility of shared
reservoirs, and a Josephson type tunneling~\cite{lagoudakis_coherent_2010}
between adjacent sites, could provide new phenomena previously unobserved in
driven bosonic systems.  Magnetic phase transitions, geometric frustration and
spontaneous pattern formation of spin in lattices, domain formation, topological
spin insulators, and topological defects are a few examples of magnetic systems
that could be studied, all within a highly-controlled bosonic
many-body system.

While the condensed polariton lattice resembles nanomagnet
arrays~\cite{jamet_magnetic_2001,hai_long_2010}, it has the inherent advantages
of tunable nonlinearity, longer spin relaxation time, ps response, rapid
optical addressing and manipulation, and adaptable scalability. The
spin-polarized state at zero magnetic field is retained for many seconds, 10
orders of magnitude longer than the condensation time, making it a suitable
candidate for optical spin-based memories.

Spin switching with only a fraction of the condensate density, which is a
direct result of the nonlinearity in our system, can be used for low power
optical and electrical sensing and spin switches. Finally, the possibility of
coherent driving allows the realization of superpositions of spin-up and
spin-down states, which are the key requirements for quantum information
processing.

\section*{Acknowledgments}
We acknowledge discussions with B. L. Altshuler and N. G. Berloff. This work was supported by EPSRC Grant No. EP/G060649/1, EP/L027151/1, EU Grant No. CLERMONT4 235114, EU Grant No. INDEX 289968, ERC Grant No. LINASS 320503, the Leverhulme Trust Grant No. VP1-2013-011, Spanish MEC (Grant No. MAT2008-01555), the Greek GSRT ARISTEIA Apollo program and Fundación La Caixa, and Mexican CONACYT Grant No. 251808. F. P. acknowledges financial support from EPSRC at the University of Cambridge and a Schrödinger grant at the University of Oxford. The data corresponding to all the figures in this paper can be found at \url{https://www.repository.cam.ac.uk/handle/1810/248036}

.

\appendix
\section{Experimental methods\label{app:expmethods}}
The cavity's top (bottom) distributed Bragg
reflector (DBR) is made of 32 (35) pairs of Al$_{0.15}$Ga$_{0.85}$As/AlAs layers
of \SI{57.2}{\nm}/\SI{65.4}{\nm}. Four sets of three \SI{10}{nm} GaAs quantum
wells (QW) separated by \SI{10}{nm} thick layers of Al$_{0.3}$Ga$_{0.7}$As are
placed at the maxima of the cavity light field.  The 5$\lambda$/2
(\SI{583}{\nm}) cavity is made of Al$_{0.3}$Ga$_{0.7}$As. The microcavity
sample is chemically etched from the substrate side to form \SI{300}{\um}
diameter membranes allowing optical access from the back of the sample for
resonant excitation [FIG.~\ref{fig:1}(a)]. The sample shows condensation under nonresonant
excitation~\cite{tsotsis_lasing_2012}.  The excitation laser is a single-mode
CW Ti:Sapphire, which is amplitude modulated using an AOM with a rise time of \SI{100}{\ns}. To pattern the pump intensity a spatial light modulator was used~\cite{cristofolini_optical_2013}.

\section{Strain-induced linear polarization splitting\label{app:polsplit}}

\begin{figure*}
	\centering 
	\includegraphics[width=0.80
    \textwidth]{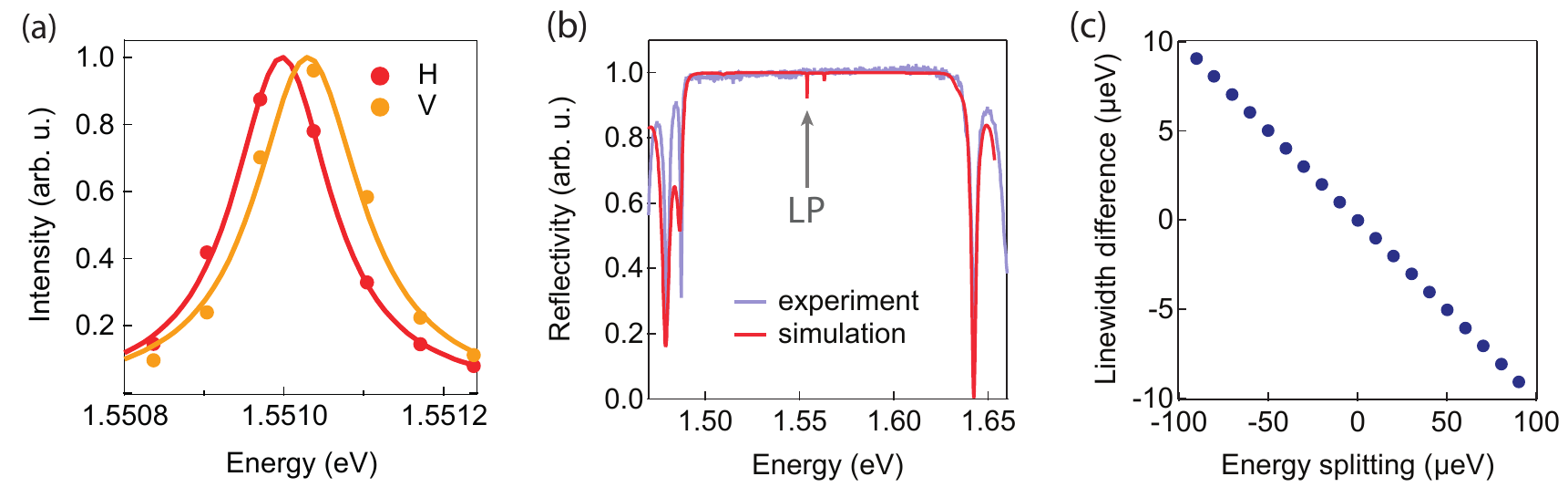} \caption{(a) Energy-resolved emission of the ground state polariton
        for horizontal (H) and vertical (V) polarizations, 
         split in energy by $\SI{\sim 30}{\ueV}$
         (b) Transfer matrix calculations (red)
        and experimental reflectivity of sample. The lower polariton
        mode (marked by an arrow), is offset by \SI{10}{\milli\electronvolt}
        below the center of the stopband. (c) Transfer matrix calculations
        showing the linewidth difference versus the energy splitting. The
        linewidth difference is $10\%$ of the energy splitting.
    } \label{fig:transmat} 
\end{figure*}

The measured energy of the ground state $X$- and $Y$-polarized photoluminescence
far below threshold is shown in FIG.~\ref{fig:transmat}(a).
The energy splitting of the linearly polarized modes varies across the sample
surface, reaching its maximum at the edges and minimum at the center of the
membranes. This observation suggests that the splitting is correlated to the
level of strain across the microcavity structure, since the latter is expected
to possess a similar spatial dependence: namely to be strongest at the boundary
between etched and non-etched regions and relax towards the central parts of
the membranes. Note that we observe all the same phenomena even in unetched
samples, as strain is universally present from the III-V heterostructure
growth.

Strain-induced splitting of the initially degenerate polariton states at $k=0$
into orthogonal linearly polarized modes has been demonstrated in previous
works~\cite{balili_huge_2010,klopotowski_optical_2006}. Both the excitonic and
the photonic parts of the polariton can be affected by strain to produce such an
anisotropy. In the former case, the initial splitting of the bright exciton
states due to exchange interactions are enhanced by strain-induced mixing of the
heavy and light hole valence bands, thereby reducing the symmetry of the
QW~\cite{aleiner_anisotropic_1992,ivchenko_heavy-light_1996,balili_huge_2010}.
In the latter case, strain induces a small birefringence in the cavity and/or
DBRs, hence lifting the degeneracy between the $[110]$ and $[1\bar{1}0]$
axes~\cite{klopotowski_optical_2006}. 

Because of the finite curvature of the cavity stopband, splitting the lower
polariton into two orthogonal modes necessarily induces a difference of their
linewidths as well. For the microcavity structures studied
in this work the polariton modes are located on the low-energy side of the
stopband [FIG.~\ref{fig:transmat}(b)]. In this case the mode possessing higher energy will exhibit a narrower linewidth,
since it is located closer to the stopband center, where the DBR reflectivity is
at its maximum. Correspondingly, the lower energy mode will exhibit a larger
linewidth, since it is located closer to the edge of the stopband, where the DBR
reflectivity starts to drop. For an energy splitting of the order of
\SI{30}{\ueV}, transfer matrix calculations predict a linewidth
difference of approximately \SI{3}{\ueV} [FIG.~\ref{fig:transmat}(c)].

\begin{figure}
	\centering 
	\includegraphics[width=0.48
    \textwidth]{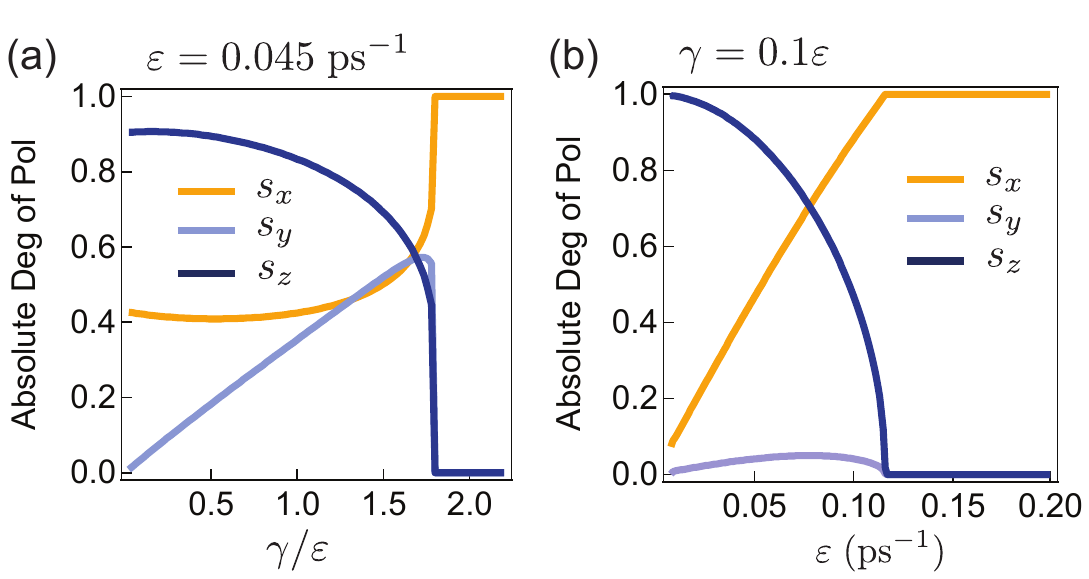} \caption{(a) Absolute degree of polarization vs
        $\gamma$ for a fixed energy splitting of
        $\varepsilon=\SI{0.045}{\per\ps}$. (b) Absolute degree of polarization
        vs energy splitting $\varepsilon$, for the case when
        $\gamma=0.1 \varepsilon$.
    } \label{fig:dEgamma} 
\end{figure}

This small difference in linewidth is not resolvable with our instruments. 
However, it should be noted that any nonzero linewidth difference, as long as it is
negative sign relative to the energy splitting, eventually leads to the
bifurcation of the circular polarization at a critical threshold
[FIG.~\ref{fig:dEgamma}(a)]. Moreover, if the ratio of the linewidth difference
to energy splitting is kept constant (as we have $\gamma=0.1\varepsilon$
according to our transfer matrix simulations), the condensate becomes linearly
polarized at high energy splittings [FIG.~\ref{fig:dEgamma}(b)]. This is indeed
what we observe at the edges of the membrane, where the energy splitting is as
high as $\SI{100}{\ueV}$.

\section{Variable cross spin saturation\label{app:varcross}}
\makeatletter{}\vspace{0.5\baselineskip}\noindent For the case when the same-spin and cross-spin gain-saturation nonlinearities are different we have:
\begin{subequations}\label{Evol2psi2}
 \begin{alignat}{2}
   \begin{split}
 \dot{\psi}_{+1}=&\half\left[w-\half(\mu|\psi_{+1}|^2
                  +(2\eta-\mu)|\psi_{-1}|^2)\right]\psi_{+1}\\
                 &-\half(\gamma-\mi\ve)\psi_{-1}
                 -\ihalf\left[\alpha_1|\psi_{+1}|^2+\alpha_2|\psi_{-1}|^2\right]\psi_{+1},
    \end{split}\\
  \begin{split}
 \dot{\psi}_{-1}=&\half\left[w-\half(\mu|\psi_{-1}|^2
                  +(2\eta-\mu)|\psi_{+1}|^2)\right]\psi_{-1}\\
                 &-\half(\gamma-\mi\ve)\psi_{+1}
                 -\ihalf\left[\alpha_1|\psi_{-1}|^2+\alpha_2|\psi_{+1}|^2\right]\psi_{-1},
 \end{split}
\end{alignat}
\end{subequations}
where $w=W-\Gamma$ and we have now two saturation parameters, $\eta$ and $\mu$. The saturation is
controlled by individual occupations of circular components when $\mu=2\eta$, and the saturation is
controlled by the total occupation when $\mu=\eta$. In general, $\eta\leqslant\mu\leqslant2\eta$.
Equations above in the matrix form read
\begin{align}\label{EvolPsi}
  \begin{split}
 \frac{d\Psi}{dt}=&\frac{1}{2}\left[w - \eta S - (\mu-\eta) S_z\sigma_z\right]\Psi
 -\frac{1}{2}(\gamma-\mi\varepsilon)\sigma_x\Psi\\
 &-\frac{\mathrm{i}}{2}\left[(\alpha_1+\alpha_2)S+(\alpha_1-\alpha_2)S_z\sigma_z\right]\Psi,
   \end{split}
\end{align}
and the equations for pseudospin components are
\begin{subequations}\label{EvolSpin2}
\begin{eqnarray}
  && \dot{S}_x =(w-\eta S)S_x - \gamma S - \alpha S_z S_y, \\
  && \dot{S}_y =(w-\eta S)S_y + \ve S_z  + \alpha S_z S_x, \\
  && \dot{S}_z =(w-\mu  S)S_z - \ve S_y.
\end{eqnarray}
\end{subequations}

\subsection{Linearly polarized condensates (parity conserved)}

These particular solutions are given by $S_y=S_z=0$, $S_x=\pm S$, and $S=(w\mp\gamma)/\eta$.
$X$-polarized condensates (upper sign) can exist for $W>\Gamma+\gamma$. $Y$-polarized states (lower
sign) appear for $W>\Gamma-\gamma$. Considering their stability with respect to small
fluctuations: $S_y=y$, $S_z=z$, $S_x=\pm S_0+x$, and $S=S_0\mp x$, with $S_0=(w\mp\gamma)/\eta$, we
have linearized equations
\begin{subequations}\label{StabEqs}
\begin{eqnarray}
  && \dot{x}=-\eta S_0 x, \\
  && \dot{y}=\pm\gamma y+(\ve\pm\alpha S_0)z , \\
  && \dot{z}=(\pm\gamma -(\mu-\eta)S_0)z - \ve y .
\end{eqnarray}
\end{subequations}
Taking $x,y,z\propto e^{\lambda t}$ we see that fluctuations in $x$ always decay, while
(\ref{StabEqs}b,c) produce the equation for the Lyapunov exponent $\lambda$,
\begin{equation}\label{Lyapun}
    (\lambda\mp\gamma)^2+(\mu-\eta)S_0(\lambda\mp\gamma)+\ve(\ve\pm\alpha S)=0.
\end{equation}

The stability of the $Y$-polarized condensate is lost when one root of this equation crosses zero. This
corresponds to the critical occupation $S_c$ and critical pumping $W_2$,
\begin{equation}\label{W22}
    S_c=\frac{\gamma^2+\ve^2}{[\alpha\ve-(\mu-\eta)\gamma]}, \qquad
    W_2=\Gamma-\gamma + \eta S_c.
\end{equation}

The stability of $X$ polarized state depends on the interrelation between $\gamma$ and
$(\mu-\eta)S_0$. For $\mu=\eta$ this state is always unstable. However, if $\mu>\eta$ the stable $X$
polarized state can be formed for large enough condensate occupations (far above the threshold).

\subsection{Elliptically polarized condensates (parity broken)}

These solutions with $S_z\ne 0$ are given by
\begin{subequations}\label{WLStates2}
\begin{eqnarray}
 &&S_x=-\frac{1}{\alpha\ve}\left[\ve^2+(w-\eta S)(w-\mu S)\right], \qquad\\
 &&S_y=\frac{(w-\eta S)}{\ve}S_z, \qquad\\
 &&S_z=\pm\ve\sqrt{\frac{S^2-S_x^2}{\ve^2+(w-\mu S)^2}}.
\end{eqnarray}
\end{subequations}
Substitution into (\ref{EvolSpin2}a) gives
\begin{align}\label{EqForS}
\begin{split}
 &(\mu-\eta)\ve\left[\ve^2+(w-\eta S)(w-\mu S)\right]\\
 &+\alpha^2\ve(w-\mu S)S+\alpha\gamma\left[\ve^2+(w-\mu S)^2\right]=0.
\end{split}
\end{align}
The positive root of this equation for $S$ should be taken. Also, it is necessary to satisfy the
condition $|S_x|\leqslant S$, which gives $w\geqslant(W_2-\gamma)=w_c$. This means that the weak
lasing solutions appear continuously from the $Y$-polarized solution at the critical pumping $W_2$.

The stability of weak lasing states can also be lost. Numerical analysis shows the following
typical scenario of evolution of the condensate polarization state with increasing $w$ for
$\mu>\eta$. First, the $Y$ linearly polarized state is formed. Then it transforms into a weak lasing,
elliptically polarized state. The stability of the latter is also lost with increasing $w$,
resulting in some irregular, quasi-chaotic dynamics and/or in the oscillatory motion of the pseudospin
vector. Finally, for large $w$ the stable $X$ polarized state is formed.

\section{Estimate for particle number\label{sec:particlenumber}}

The condensate particle number is experimentally measured by:
\begin{equation}\label{eq:particlen}
n=\frac{\Phi \tau}{ \vert C \vert^2},
\end{equation}
where $\Phi$ is the photon flux, $\tau=\SI{10}{\ps}$ is the polariton lifetime, and $\vert C\vert^2=0.4$ is the photon Hopfield coefficient. The photon flux is measured by:
\begin{equation}
\Phi=\frac{\alpha R}{\eta}, 
\end{equation}
where $\alpha=3.5\;\mathrm{e^-/count}$ is the photoelectron sensitivity of the
CCD, $\eta=0.0021$ is the total detection efficiency including the camera
quantum efficiency and the total optical transmission efficiencies, and
$R=\SI{1.9E10}{\per\second}$ is the spatially-integrated count rate of the CCD.
Inserting these values in Eq.~(\ref{eq:particlen}) gives the particle number
$n\simeq800$, at $P=1.7 P_{\text{th}}$.

\section{Resonance width\label{app:reswidth}}

\begin{figure}[htb]
	\centering 
	\includegraphics[width=0.48
	\textwidth]{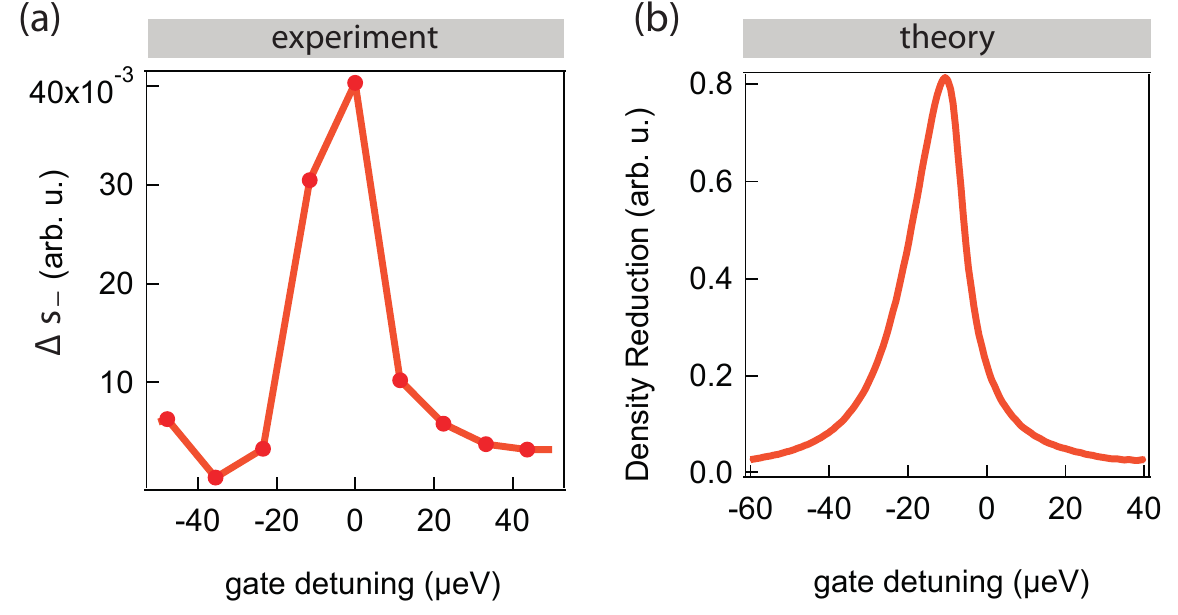} \caption{
   (a) The reduction of the spin-down intensity versus the detuning of the gate laser frequency is shown. A sharp resonance is visible when the laser is in resonance with the condensate. (b) The numerical calculations show a similar resonance effect, with the center of the peak slightly redshifted due to the linear polarization energy splitting $\ve$.
   } \label{fig:resPeak} 
\end{figure}

We study the reduction of the opposite component of the condensate circular
polarization to the gate laser [marked by $\Delta s_-$ in
FIG.~\ref{fig:res}(d)] as the `detuning' of the gate varies, while the gate
power remains constant. In order to change the detuning of the gate laser
frequency with respect to the condensate, instead of changing the frequency of
the laser, we tune the condensate frequency by changing the pump intensity. The
trapped condensate `blueshifts' as the pump intensity is increased, mainly due
to the repulsive interaction between the polaritons. This contrasts with the
case of untrapped condensates, where interactions of polaritons with the
excitons in the reservoir is the source of
blueshifts~\cite{wouters_spatial_2008,wertz_spontaneous_2010,askitopoulos_polariton_2013}.
In the 4 spot trapped geometry, we have a blueshift of \SI{6}{\ueV/\mW} and as
a result we can tune the condensate energy accurately with respect to that of
the gate.  FIG.~\ref{fig:resPeak}(a) shows $\Delta s_-$ with respect to the
detuning of the gate laser. We observe a sharp resonance, with a full width at
half maximum (FWHM) of \SI{11\pm 6}{\ueV}. FIG.~\ref{fig:resPeak}(b) shows the
theoretical curve versus the gate detuning ($\hbar\omega_g$), which resembles a
Lorentzian profile with a linewidth of \SI{17}{\ueV}. Theoretically, the
linewidth of the resonance is defined by the noise in Eqs.~\eqref{eq:noise}.
The FWHM of the Lorentzian resonance peak is $D/(1-s_z^2)n$. We note that the
resonance frequency $\omega_c$ is slightly redshifted, due to the effect of the
linear polarization splitting and the decay rate splitting, \begin{equation}
\omega_c=\frac{1}{2}\left[ (\alpha_1+\alpha_2) S - \frac{(\gamma S_y S_z + \varepsilon S
S_x)}{(S_x^2+S_y^2)} \right].  \end{equation}

\section{The Kramers flip rate\label{app:kramers}}
\makeatletter{}From Eq.~\eqref{eq:noise} we can obtain the spin vector equations:
\begin{subequations}\label{eq:LEqsS}
 \begin{eqnarray}
 &&\dot{S}_x= - g(S) S_x - \gamma S - \alpha S_z S_y + F_x(t), \\
 &&\dot{S}_y= - g(S) S_y + \ve S_z + \alpha S_z S_x + F_y(t), \\
 &&\dot{S}_z= - g(S) S_z - \ve S_y + F_z(t),
 \end{eqnarray}
\end{subequations}
where the correlators of real-function noise $F_i(t)$ are
\begin{equation}
 \label{NoiseS}
 \left<F_i(t)F_j(t^\prime)\right>=2DS\delta_{ij}\delta(t-t^\prime), \qquad i,j=x,y,z.
\end{equation}

Here we consider limiting case when two parity breaking states are formed near the north and south
poles of the Poincar\'e sphere, i.e., $|S_z|\gg|S_{x,y}|$. The spin components of fixed states in
this limit are $S_{x0}\simeq -\ve/\alpha$, $S_{y0}\simeq -\gamma/\alpha$, $S_z\simeq\pm S_0$, where
$S_0$ is the root of $S=\gamma\ve/\alpha g(S)$. In what follows we also denote $g_0=g(S_0)$ and by
assumption $g_0\ll\ve,\gamma$. Being excited away from the fixed state, the spin exhibits fast
self-induced Larmor precession and slow relaxation. The precession frequency is $\omega=\alpha
S_z$, so that $\omega=\gamma\ve/g_0$ near the stationary states.

The spin should be driven by noise into the equatorial plane ($S_z=0$) in order to flip. This
Kramers problem can be simplified if we perform averaging over fast precession of the spin vector.
This treatment is valid as long as $|S_z|\gg|S_{x,y}|$. Consider the motion in the north
hemisphere. We assume that the number of polaritons does not change during the flip, i.e., $S$ is
fixed to $S_0$, and we consider the case of large occupations, $\ln S_0\gg 1$. Omitting the
$g$-terms from Eqs.\ (\ref{eq:LEqsS}a,b) we find that for a given value of $S_z>0$ the averages
over one cycle of the other two components are $\left<S_x\right>=-\ve/\alpha$ and
$\left<S_y\right>=-\gamma S_0/\alpha S_z$. Then, from (\ref{eq:LEqsS}c) we obtain the equation for
slow evolution of $S_z$
\begin{subequations}
\label{SlowSz}
\begin{equation}
  \frac{dS_z}{dt} = - g_0 S_z + g_0 \frac{S_0^2}{S_z} + F_z(t)
  =-\frac{dU(S_z)}{dS_z}+F_z(t),
\end{equation}
\begin{equation}
  U(S_z)=g_0\left( \half S_z^2-S_0^2\ln S_z \right).
\end{equation}
\end{subequations}

This expression for the effective potential $U(S_z)$ is not valid for small $S_z$, where it
diverges logarithmically. The top of the barrier should be cut off when $S_z$ becomes comparable to
$|\left<S_y\right>|$, that is for $S_z\simeq S_0\sqrt{g_0/\ve}$. The value of potential on the top
of the barrier is then $U_b\simeq -g_0 S_0^2\ln(S_0\sqrt{g_0/\ve})$. The bottom of the well is
positioned at $U_0=g_0 S_0^2(\half-\ln S_0)$. As a result, the spin should overcome the barrier
\begin{align}\label{DeltaU}
  \Delta U=U_b-U_0&=\frac{1}{2} g_0 S_0^2\left[\ln\left(\frac{\ve}{g_0}\right)-1\right]
  \nonumber \\
  &\simeq \frac{1}{2} g_0 S_0^2 \ln\left(\frac{\ve}{g_0}\right).
\end{align}
Using the known result for the Kramers first passage time in the one-dimensional
problem~\cite{zwanzig_nonequilibrium_2001}, we obtain the spin-flip rate
\begin{align}
\label{KramersR}
  R_K=\frac{R_0}{2\pi}\exp\left\{-\frac{\Delta U}{DS_0} \right\}
  &=\frac{R_0}{2\pi}\exp\left\{-\frac{g_0 S_0}{2D} \ln\left(\frac{\ve}{g_0}\right) \right\},\\ \nonumber
  S_0&=\frac{\gamma\ve}{\alpha g_0}.
\end{align}
This expression assumes $g_0 S_0/D\gg 1$ and $\ve\gg g_0$. The pre-exponent $R_0$ cannot be written
exactly by this method, since we do not know the shape of the effective potential near the top of
the barrier. It can be estimated as $R_0\simeq\sqrt{\ve g_0}$.

\section{Simulations and numerical parameters\label{app:simparam}}
The parameters used for all the simulations are as follows:\vspace{1em} \\
0D simulations: $\eta=\SI{.01}{\ps^{-1}}$; $\Gamma=\SI{.1}{\ps^{-1}}$;
$\hbar\varepsilon=\SI{30}{\ueV}$; $\gamma=0.1\varepsilon$; $\hbar\alpha_1=\SI{10}{\ueV}$; $\alpha_2=-0.5\alpha_1$\vspace{1em}\\
2D simulations: $\hbar\alpha_1=\SI{3}{\ueV\um^2}$; $\alpha_2=-0.5\alpha_1$; $\hbar g_r=\SI{46}{\ueV\um^2}$; $g_P=g_R/4$; $\Lambda=0.1$;
$m^*=5.1\times10^{-5}m_e$; $\gamma_R=\SI{10}{\ps^{-1}\um^2}$;
$\hbar\varepsilon=\SI{7}{\ueV}$; $\gamma=0.2 \varepsilon$.

\bibliography{main}

\end{document}